\documentclass[useAMS,usenatbib]{mn2e}
\usepackage{graphicx}
\usepackage{lastpage}

\title[The Mg~II UV feature in TTSs]{Constraints to the magnetospheric properties of T Tauri stars - II.
The Mg~II ultraviolet feature}

\author[L\'opez-Mart\'inez, F. and G\'omez de Castro, A.I.]{Fatima L\'opez-Mart\'{i}nez$^{1, 2}$ and Ana In\'es G\'omez de Castro$^1$ \thanks{E-mail: aig@ucm.es}\\
$^1$AEGORA Research Group, Universidad Complutense de Madrid, Plaza de Ciencias 3,
28040 Madrid, Spain \\
$^2$ Isaac Newton Group of Telescopes, Apto. 321, E-38700, Santa Cruz de la Palma, Canary Islands, Spain}

\begin{document}
\date{Accepted 2014 December 17. Received 2014 December 15; in original form 2014 July 11}

\pagerange{\pageref{firstpage}--\pageref{lastpage}} \pubyear{}

\maketitle

\label{firstpage}

\begin{abstract}
The atmospheric structure of T Tauri Stars (TTSs) and its connection with the large
scale outflow is poorly known. Neither the effect of the magnetically mediated interaction between the star and the disc in the stellar atmosphere is well understood. The Mg~II multiplet is a fundamental tracer of TTSs atmospheres and outflows, and is the strongest feature in the near-ultraviolet spectrum of TTSs. 
The \textit{International Ultraviolet Explorer} and \textit{Hubble Space Telescope} data archives provide a unique set to study the main physical compounds contributing to the line profile and to derive the properties of the line formation region. The Mg~II profiles of 44 TTSs with resolution 13,000 to 30,000 are available in these archives. In this work, we use this data set to measure the main observables:
flux, broadening, asymmetry, terminal velocity of the outflow, and the velocity of the Discrete Absorption Components.
For some few sources repeated observations are available and variability has been studied.
There is a warm wind that at sub-AU scales absorbs the blue wing of the Mg~II profiles.
The main result found in this work is the correlation between the line broadening, Mg~II flux, terminal velocity of the flow and accretion rate. 
Both outflow and magnetospheric plasma contribute to the Mg~II flux.
The flux-flux correlation between Mg~II and C~IV or He~II is confirmed; however, no correlation is found between the Mg~II flux and the ultraviolet continuum or the H$_2$ emission.
\end{abstract}


\section{Introduction}

T Tauri stars (TTSs) are late type, Pre-Main Sequence (PMS) stars with masses below $\sim 2 M_{\odot}$. Classical T Tauri Stars (CTTSs) are roughly solar-mass stars that are accreting gas from their circumstellar discs, whereas Weak line T Tauri Stars (WTTSs) have negligible accretion rates
\citep[see][for a recent review]{aig2013a}.

The detection of rotationally modulated emission from hot ($\sim 10,000$~K) plasma both in the optical range
\citep{bouvier1990} and in the ultraviolet \citep{simon1990,aig1996} pointed out that matter in-fall  is not occurring over all the stellar surface but rather it is channelled by the stellar magnetic field. TTSs photospheric magnetic fields are $\sim 1$~kG \citep[see][for a recent compilation]{johnskrull2007}. Assuming that T Tauri magnetospheres are predominantly bipolar on the large scale, \citet{camenzind1990} and \citet{koenigl1991} showed that the inner accretion disc is expected to be truncated by the magnetosphere at a distance of a few stellar radii above the stellar surface for typical mass accretion rates of $10^{-9}$ to $10^{-7} M_{\sun} yr^{-1}$  \citep{basri1989,hartigan1995,gullbring1998}. Disc material falls from the inner disc edge onto the star along the magnetic field lines, giving rise to the formation of magnetospheric accretion columns. As the free falling material in the funnel flow eventually hits the stellar surface, accretion shocks develop near the magnetic poles \citep[see for example][]{romanova2012}.

The ultraviolet (UV) luminosities of the TTSs exceed by 1-2 orders of magnitude those observed in main sequence stars of the same spectral types. This excess is associated with the accretion process that transports material onto the stellar surface enhancing the flux radiated by magnetospheric/atmospheric tracers, typically the ultraviolet (UV) resonance multiplets of N~V, C~IV, Si~IV, He~II, C~III, C~II, Si~II, Fe~II, Mg~II, Ly-$\alpha$ and O~I \citep[see][for a recent review of the UV properties of TTSs]{aig2009a}. Though the UV excess of TTSs is well known since the early 80's, it is still unclear which is the dominant physical mechanism involved in its generation. There are evidences of it being produced in extended magnetospheres \citep{hartmann1994,hartmann1998,calvet1998,aig2012,ardila2013, aig2013b}, in accretion shocks \citep{aig1999,calvet2000,ardila2000,gullbring2000, ardila2013, aig2013b} and in outflows \citep{penston1983,calvet1985,hartmann1990,aig2001,coffey2007}.

High resolution spectroscopy is an invaluable tool to get insight into the physics associated with the release of gravitational energy in the accretion process. Emission from jets, discrete absorption components (DACs) from accreting cloudlet or episodic ejections and the radiation from the large scale excitation of the magnetosphere by the infalling gas, can be best disentangled by their kinematical signature.  The Mg~II resonance multiplet UV1  is the strongest line in the UV spectrum of TTSs, only surpassed by the Ly-$\alpha$ line that it is often strongly absorbed by the circumstellar material. The extended neutral outflow absorbs strongly the blue wind of the Ly-$\alpha$ line. Moreover, the molecular hydrogen in the circumstellar environment absorbs Ly-$\alpha$ photons that produce the H$_2$ fluorescent emission detected in the UV \citep{herczeg2002,france2012}. Compared with Ly-$\alpha$, Mg~II has the advantage of sampling a narrower temperature range preventing the pollution from the diffuse emission/absorption from the cool extended H~I envelopes. Note that the ionization potentials of Mg~I and Mg~II are 7.65~eV and 15.03~eV, respectively; henceforth Mg~II is a tracer of plasmas in the temperature range from some few thousand Kelvin up to $\sim 20,000$~K.  Moreover, the Mg~II[uv1] electronic levels distribution permits to treat the  ion as a two levels specie allowing a simple treatment of the radiation transfer \citep[see][]{catala1986}. 

From the observational point of view, the Mg~II lines have the advantage of their high Signal-to-Noise Ratio (S/N); the multiplet is observed at $\lambda \lambda 2796,2804$~\AA\ vacuum wavelength, where the sensitivity of the UV instrumentation is high, allowing to obtain high resolution profiles even with small effective area telescopes such as the \textit{International Ultraviolet Explorer (IUE)} \citep{penston1983,calvet1985,aig1998,ardila2002ii,herczeg2004}. For this reason, there is a large enough sample of observations to run a study of the TTSs as a class, including variability. The objective of this work is to run such a study.

The Mg II lines have been used by several authors to study the structure of some few TTSs and to derive different physical properties
\citep[see e.g.][]{imhoff1980,aig1997,lamzin2000}.
\citet{giampapa1981} used a sample of 13 TTSs observed with the \textit{IUE} to study the chromospheric origin of the Mg II emission and to derive the mass loss rates in the TTSs' wind. Further research on the connection between the chromosphere 
and the extended envelope was carried out by \citet{calvet1985}, who made use of simultaneous observations of the Ca~II and Mg~II lines of BP~Tau, DE~Tau, RY~Tau, T~Tau, DF~Tau, DG~Tau, DR~Tau, GM~Aur SU~Aur, RW~Aur, CO~Ori and GW~Ori to conclude that  the chromospheric structure seemed to be related with the mass of the stars. TTSs with masses above 1.5~M$_{\odot}$ seemed to produce the Mg~II emission in extended envelopes, alike the H$\alpha$ emission, while less massive TTSs have Mg~II emission produced in the chromosphere. This result was interpreted in terms of the internal structure of the star and the energy transport. Low mass, fully convective, TTSs were expected to be slower rotators. However, the authors concluded that Mg~II emission also seemed to be produced in extended regions in some low mass TTSs. 
\citet{ardila2002ii} analysed the relationship between the Mg II flux and several stellar properties using a small sample of TTSs (those observed with the Goddard High Resolution
Spectrograph (GHRS) in \textit{Hubble Space Telescope (HST)})\footnote{Based on observations made with the NASA/ESA Hubble Space Telescope, obtained from the data archive at the Space Telescope Science Institute. STScI is operated by the Association of Universities for Research in Astronomy, Inc. under NASA contract NAS 5-26555.}. The sample included: BP~Tau, T~Tau, RW~Aur, DF~Tau, DG~Tau, RU~Lup, RY~Tau, T~Tau, DR~Tau and HBC~388. However, no correlation between the Mg~II
flux and the accretion rate was found. Also, they did not find any correlation between
any parameter of the Mg~II line emission and the inclination. This result was interpreted by the authors as an evidence  of the line emission coming from a non-occulted area. However, evidence of a latitude dependent wind was claimed from the data.
Finally the comparison between the H$\alpha$ emission and the Mg~II emission of
BP~Tau, DF~Tau, RW~Aur and DR~Tau (the observations were not simultaneous) pointed out 
that the line broadening were very similar indicating that line broadening was
dominated by the kinematics of the emission region rather than by other mechanisms, i.e. stark broadening. 
The relation between the Mg~II flux and the accretion rate is at debate.
\citet{calvet2004} showed that Mg II line luminosity correlates with accretion luminosity in accreting stars and the same trend was found using 
spectra obtained with the Space Telescope Imaging Spectrograph (STIS)  
by \citet{ingleby2011}. However, this correlation was found on the basis of low resolution data and as pointed out in earlier works the Mg~II doublet is saturated. 

The Mg~II emission is a main tracer of the TTSs magnetosphere and
its study is fundamental to determine its extent and heating sources. 
In a previous work, we estimated the plasma properties in the
formation region of the semi-forbidden C~II], Fe~II]
and Si~II] lines. In that work we pointed out that these lines are formed
in the accretion flow. In three stars (DG~Tau, FU~Ori and RY~Tau)
a contribution of the outflows to the lines was observed,
suggesting that the properties in the base of the jet are similar to
those observed in the base of the accretion stream
\citep{fatima2014}.
As the 
TTSs magnetosphere is expected to end in a sheared boundary layer, acting as the magnetized interface with the Keplerian disc,
understanding the source of the Mg~II lines broadening can provide fundamental
clues on the star-disc angular momentum transport. Currently, there are 
in the \textit{IUE} and \textit{HST} archives observations of the Mg~II line profiles with resolutions
between 15,000 and 45,800 of 44 TTSs, including WTTSs,
fast and slow rotators with a large range of ages and masses. This provides a
extraordinary sample to run statistically significant tests on the properties
of the TTSs and the evolution of their magnetosphere as they approach the main sequence.
In this work, we analyse 126 observation of 44 TTSs to run such a study. For most of them,
also the Ly-$\alpha$ profile is available in the \textit{HST} archive. This information
has been used to complete the view on the circumstellar environment of the TTSs.
The Archive data are described Sect.~\ref{archivaldata}. The characteristics of the sample of TTSs 
observed by these missions are summarized in Sect.~\ref{generalproperties}. Since there are many uncertainties in the
PMS evolutionary tracks, age and masses have been derived for all
sources. The data and the measurement procedures are described in Sect.~\ref{measurements}. In Sect.~\ref{constraints},
the constraints to the physics of the TTSs outflows are analysed from the data.
The connection between accretion process and line emission is re-examined on the light of the 
new data in Sect.~\ref{discussion}. The article concludes with a summary in
Sect.~\ref{conclusions}.

\section{Archival data}
\label{archivaldata}
The Mg~II profiles were extracted from the \textit{IUE} and \textit{HST} archives
for the log of observations shown in Table~\ref{logofobservations}.
We checked the archives for all the available observations
of the Mg~II profiles for TTSs with resolutions between 13,000
and 30,000. Taking into account these characteristics, we selected
126 observations of 44 TTSs the archives.

\begin{table*}
\caption{Telescope/instrument details of the Mg~II observations of the stars in the sample. The full table is available online as Supporting Information. \label{logofobservations}}
\begin{tabular}{ccccccc}
\hline
Star & Instrument & Obs. Date & Data set & Res. &  Exposure & S/N \\
     &            & (yy-mm-dd) & Id.     &    power        & Time (s)    &     \\ \hline
AA	Tau	&	\textit{HST}/STIS	&	07-11-01	&	ob6ba7030	&	30000	&	1462.2	&	4.20	\\	\hline
AK	Sco	&	\textit{IUE}	&	86-08-06	&	LWP08847	&	13000	&	16859.8	&	7.80	\\	
		&	\textit{IUE}	&	88-04-01	&	LWP12964	&	13000	&	5099.8	&	very noisy	\\	
		&	\textit{IUE}	&	88-04-02	&	LWP12967	&	13000	&	33599.6	&	5.40	\\	
		&	\textit{IUE}	&	88-04-02	&	LWP12968	&	13000	&	9899.5	&	3.10	\\	
		&	\textit{IUE}	&	88-04-09	&	LWP13006	&	13000	&	25799.8	&	8.30	\\	
		&	\textit{HST}/STIS	&	10-08-21	&	ob6b21030	&	30000	&	1015	&	13.90	\\	\hline
BP	Tau	&	\textit{IUE}	&	81-07-24	&	LWR11130	&	13000	&	12599.6	&	8.60	\\	
		&	\textit{IUE}	&	85-10-22	&	LWP06963	&	13000	&	10799.8	&	very	noisy	\\ \hline
\end{tabular}
\end{table*}
%%%%%%%%%%%%%%%%%%%%%
The \textit{IUE} observations were obtained in high dispersion mode with R$\simeq$13,000. The Mg~II
lines were in orders 82 and 83 in the long wavelength spectrograph. The Mg~II~{\it h} 
(2796~\AA ) line is well centred in order 82 but the Mg~II~{\it k} (2803~\AA ) line
was at the edge of the orders. The echelle ripple correction introduced an enhancement
of the noise, as a result the Mg~II~{\it k} line is more noisy and shows some spikes in the
\textit{IUE} spectra. This problem was partially solved in the final \textit{IUE} data processing
both for NEWSIPS (New Spectral Image Processing System) and INES (\textit{IUE} Newly Extracted
Spectra). NEWSIPS release is accessible trough the MAST archive \citep[see][for details in the \textit{IUE} data processing]{nichols1996}.
In the INES database, also the high resolution 'concatenated' spectra are stored\footnote{The INES archive can be accessed through
http://sdc.cab.inta-csic.es/ines/index.html}. In the INES release the spectral orders are connected, eliminating the regions overlapped through a procedure designed to optimize the S/N at the edges of the orders \citep[see][for details]{cassatella2000}. A comparison between both releases can be found in \citet{gonzalez2000}. 
The data used in this article were retrieved from the INES archive. 

The \textit{HST} observations were obtained with three different instruments: GHRS, STIS 
and the Cosmic Origins Spectrograph (COS). The details on the instrument, grating, aperture, dispersion for each instrument and
configuration are provided in the log of observations for each data set.
Some of the observations obtained with STIS are included in the catalogue 
of UV stellar spectra of cool stars: CoolCat. In this case, 
the data were retrieved directly from the CoolCat web site\footnote{ 
http://casa.colorado.edu/$\sim$ayres/CoolCAT/} \citep[see][for details]{ayres2010}.
There are repeated observations of up to 17 stars in the sample, allowing to study the profiles variability. 

The Ly-$\alpha$ profiles were extracted from the \textit{HST} archive. The observations
were obtained either with STIS (E140M; G140M) or with COS (G130M) (see Table~\ref{lymanalpha}).

The lines profiles are plotted in Fig.~\ref{profiles}. The profiles are arranged in increasing
order of line broadening and asymmetry. This is also the sequence from
WTTSs to heavily accreting TTSs. In the last panel, three stars: CY~Tau, DM~Tau and
AK~Sco are plotted. These are accreting sources where the red-wards shifted wing
of the profile is more absorbed than the blue-wards shifted one. 

Most of the  Ly-$\alpha$ profiles are dominated by the geocoronal emission. In the STIS spectra,
it is observed as a narrow emission component at rest wavelength. However,
in the COS spectra, the geocoronal emission is much broader because of the wider aperture
\citep[see][for details on the Ly-$\alpha$ profiles of the TTSs obtained with COS]{france2012}.
As a result, the only relevant information about the Ly-$\alpha$ profile 
of the TTSs concerns the high velocity wings of the line. For this reason, the Ly-$\alpha$ profiles
are plotted in logarithmic scale. The absorption of Ly-$\alpha$ photons by the  
circumstellar neutral wind is readily observed; there is also Ly-$\alpha$ absorption
by the H$_2$ molecules in the circumstellar environment, that has been used to
reconstruct the underlying Ly-$\alpha$ profile \citep[see][]{herczeg2004,schindhelm2012}. 

The FUV data used in this work (C~IV, He~II and H$_2$) have been taken
from \citet{ardila2013,aig2013b,france2012}, respectively.
Most of the targets were observed with STIS (E140M) and COS (G130M; G160M).

%%%%%%%%%%%%%%%%%%%%%%%%
\begin{figure*}
\begin{tabular}{cc}
\includegraphics[width=8cm]{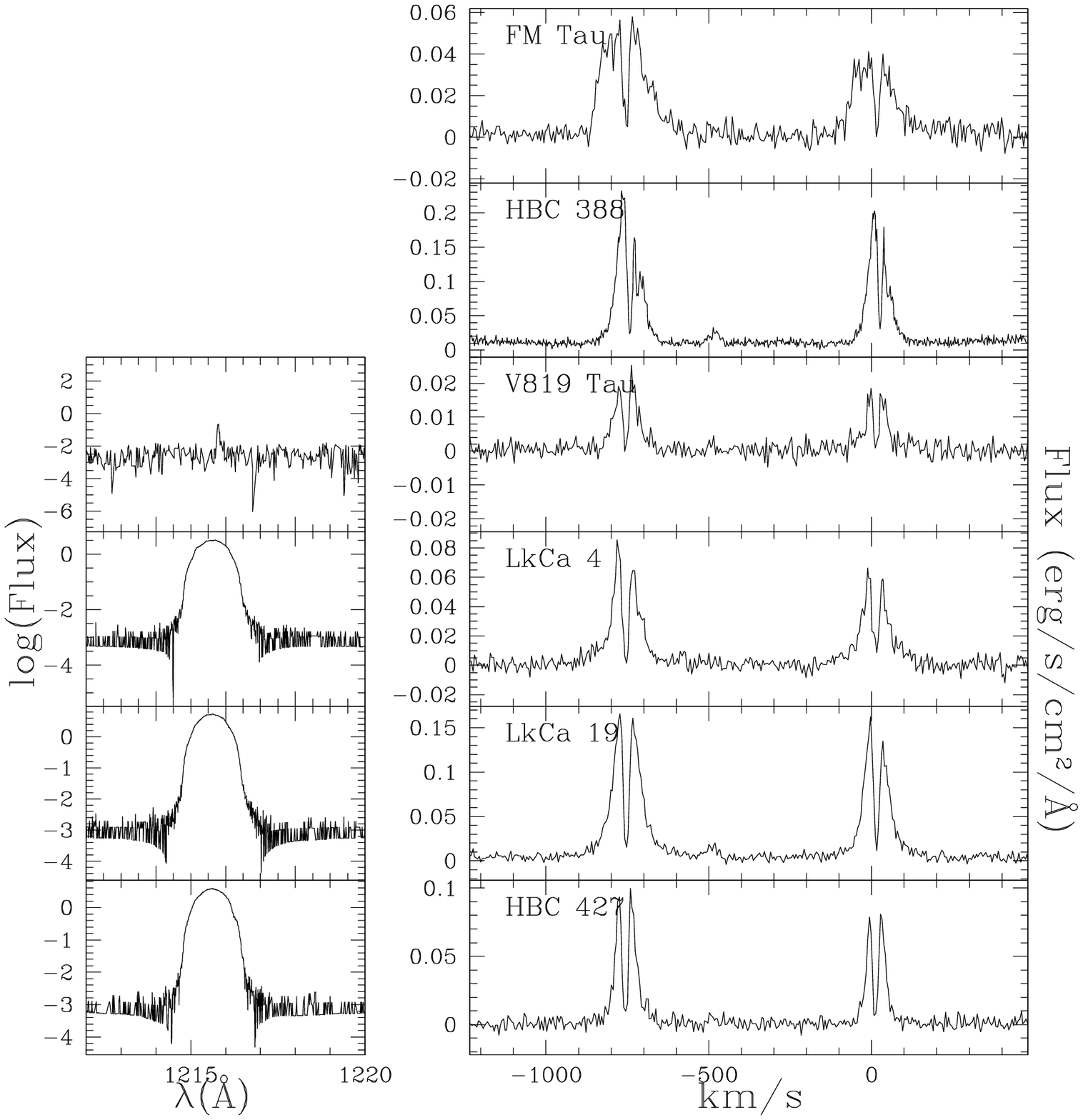} & \includegraphics[width=8cm]{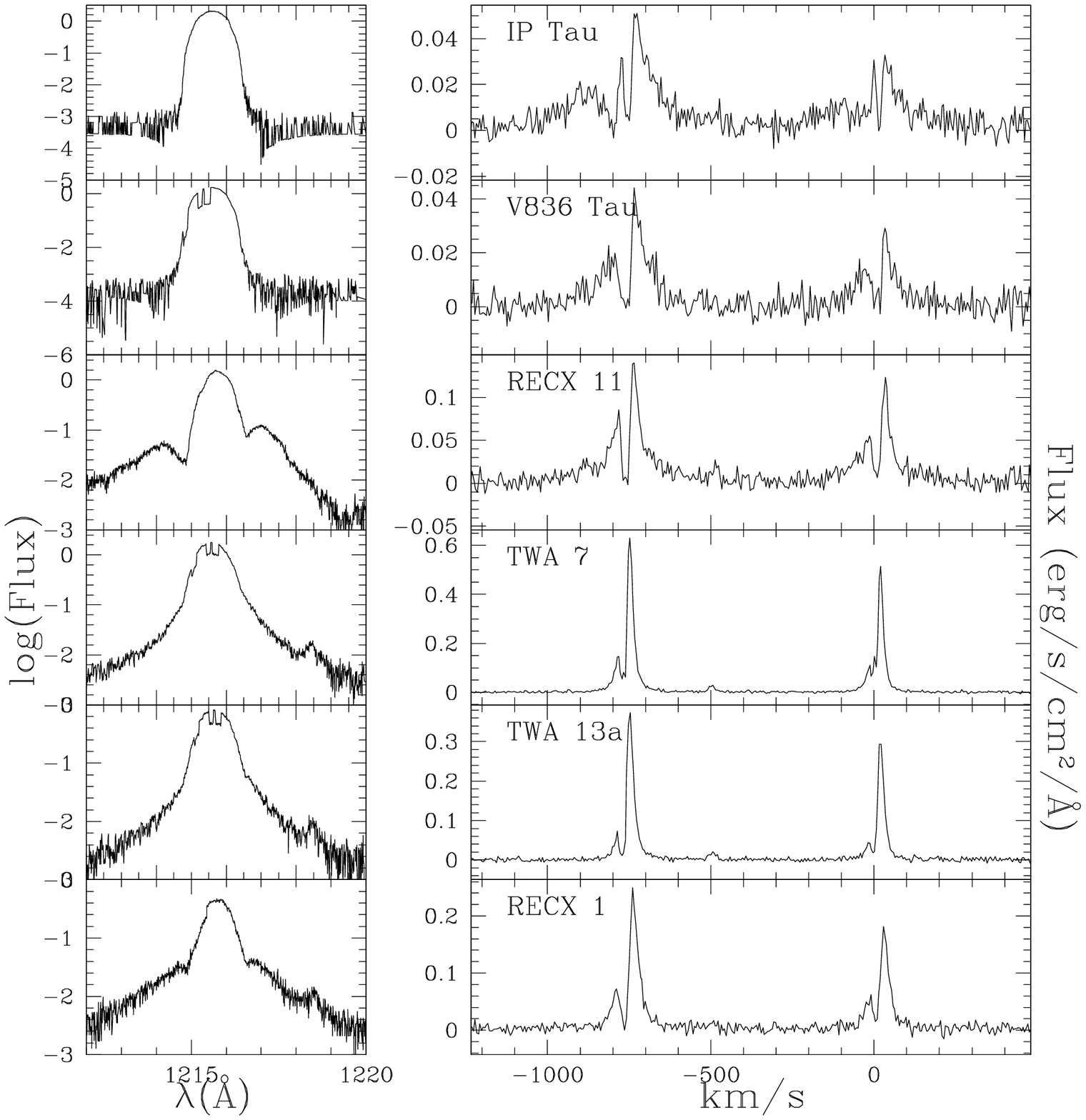} \\
\includegraphics[width=8cm]{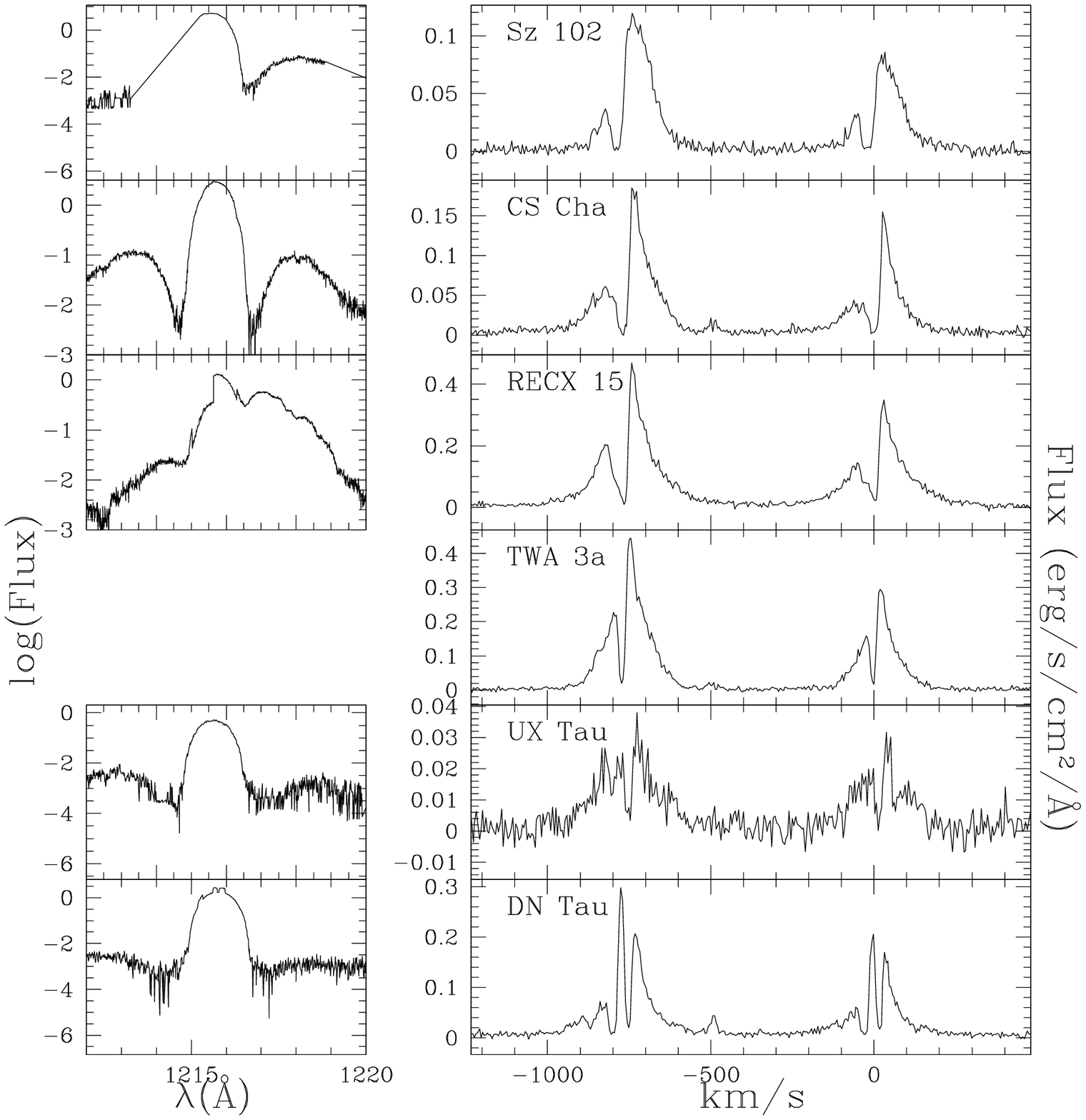} & \includegraphics[width=8cm]{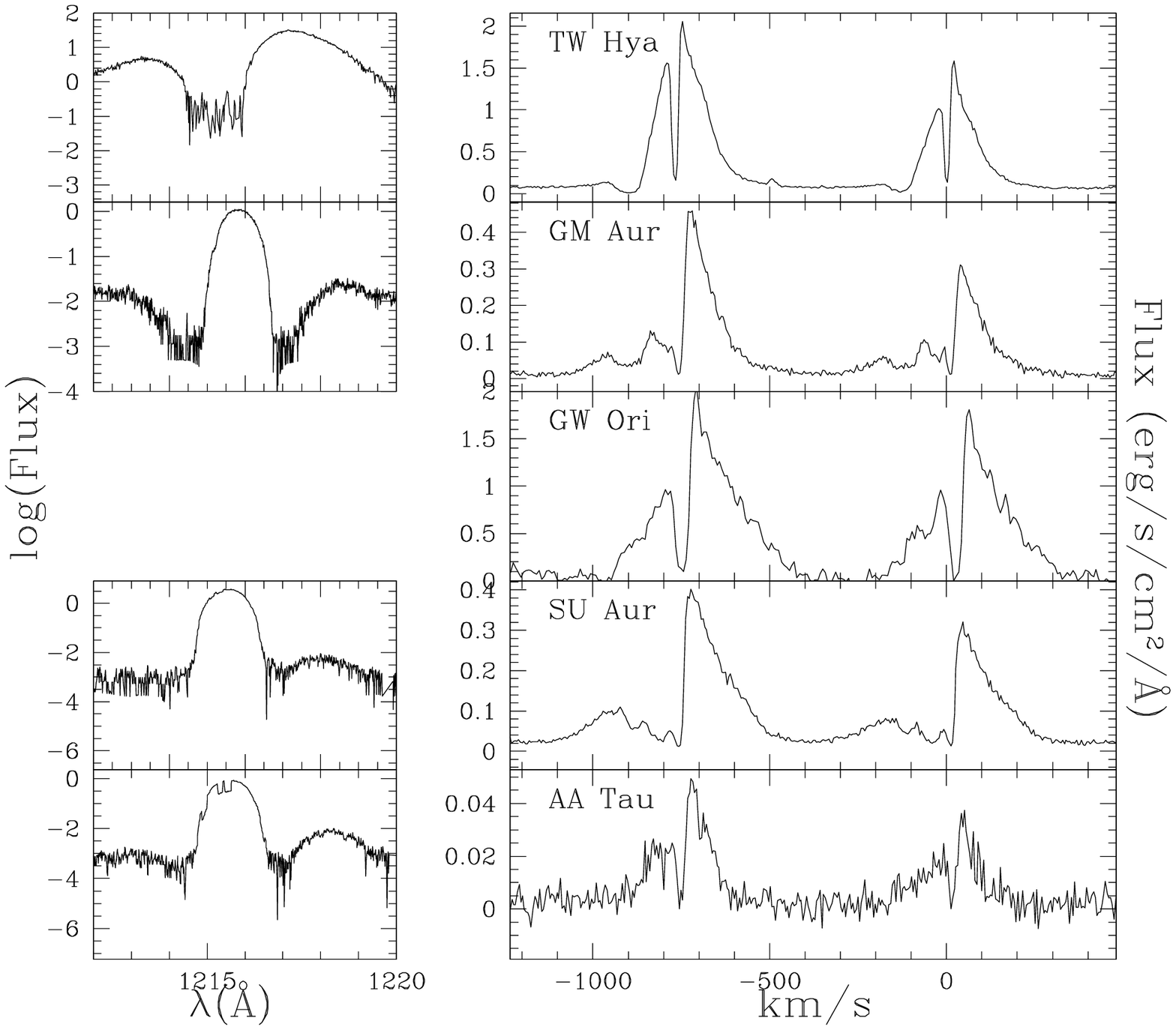} \\
\end{tabular}
\caption{Ly-$\alpha$ and Mg~II profiles of the TTSs. The Ly-$\alpha$ profiles are 
plotted in logarithmic scale to show in detail the wings since the core of the 
line is spurious geocoronal Ly-$\alpha$ emission.}
\label{profiles}
\end{figure*}
\begin{figure*}
\begin{tabular}{cc}
\includegraphics[width=8cm]{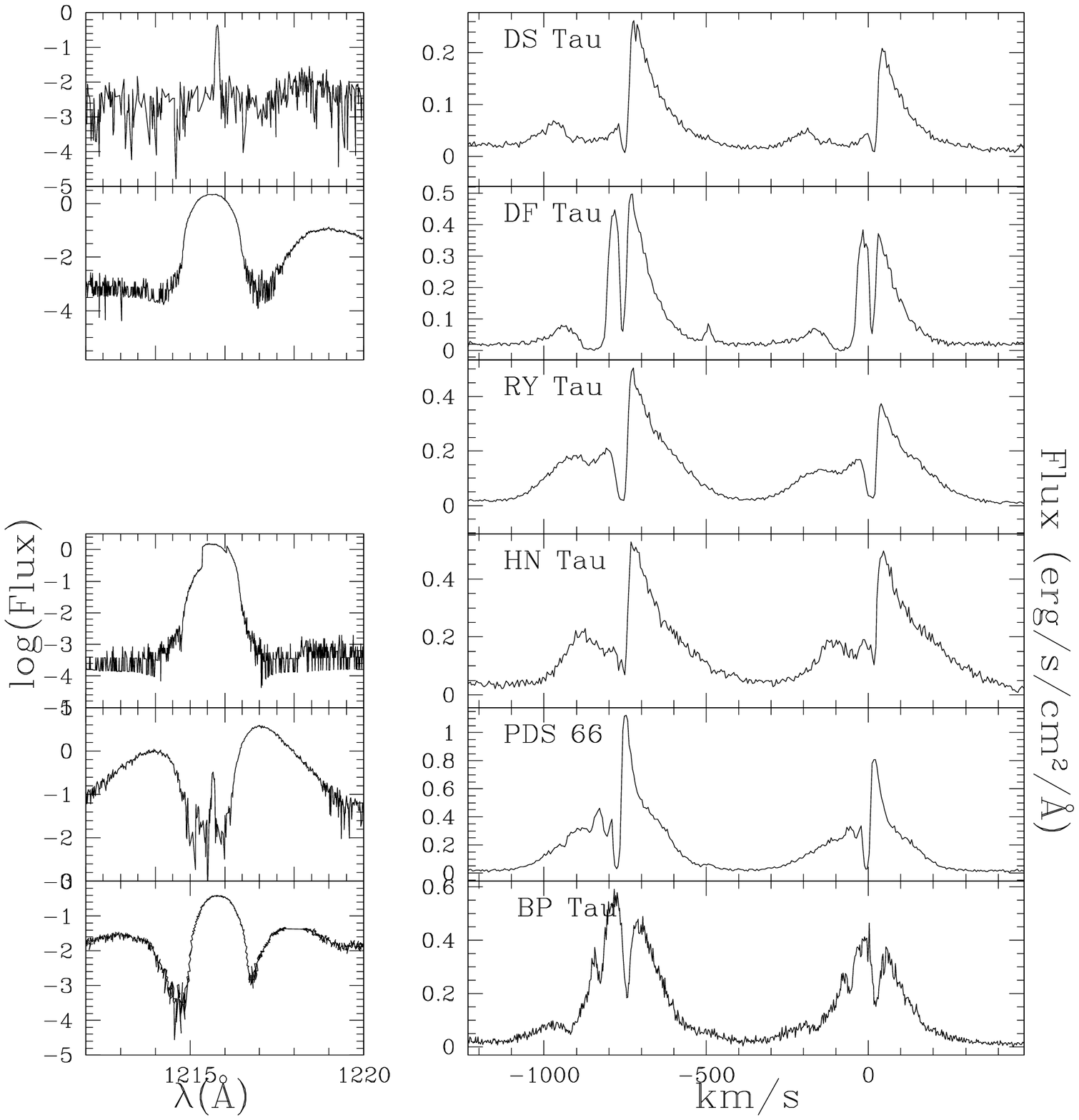} & \includegraphics[width=8cm]{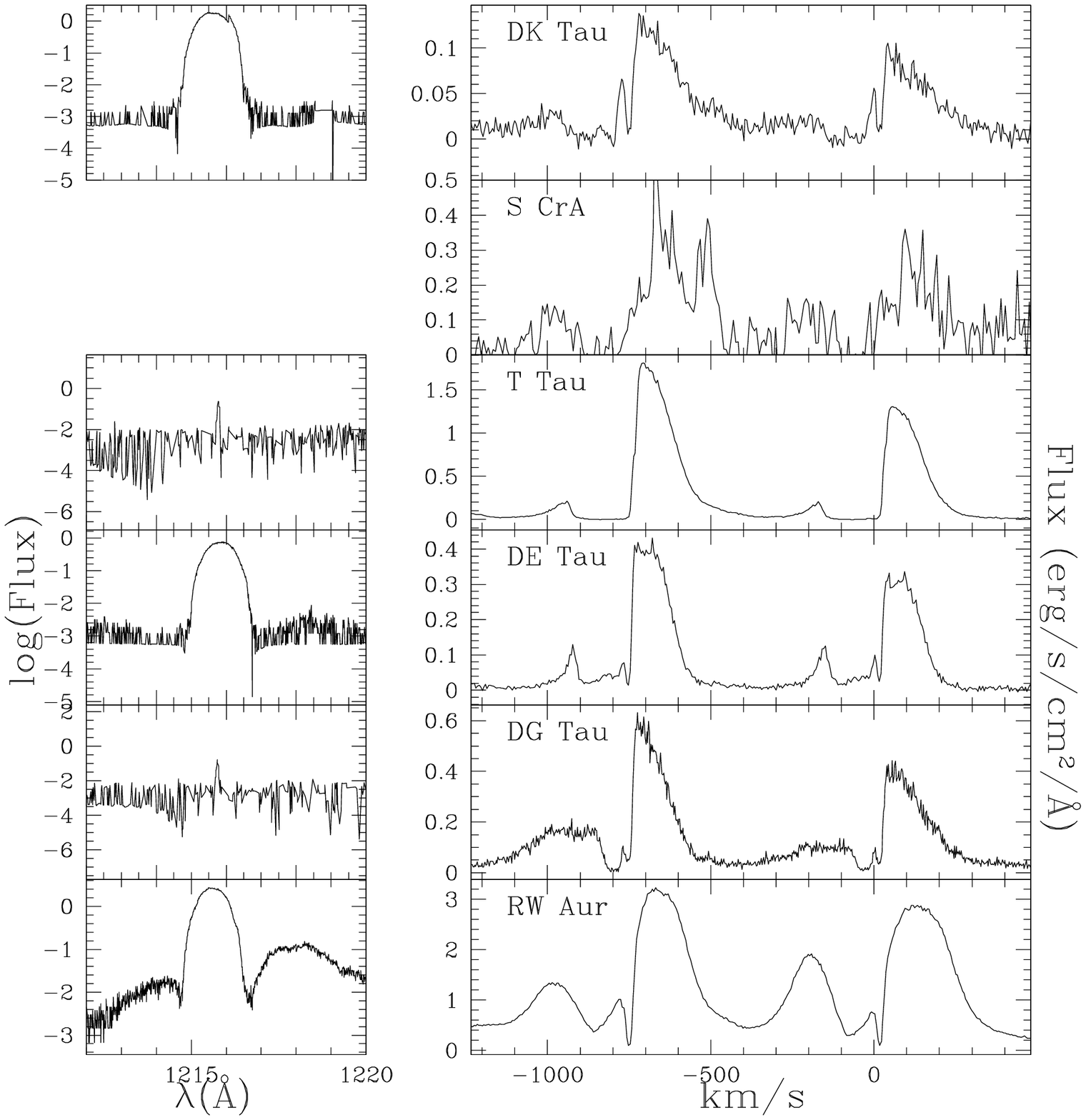} \\
\includegraphics[width=8cm]{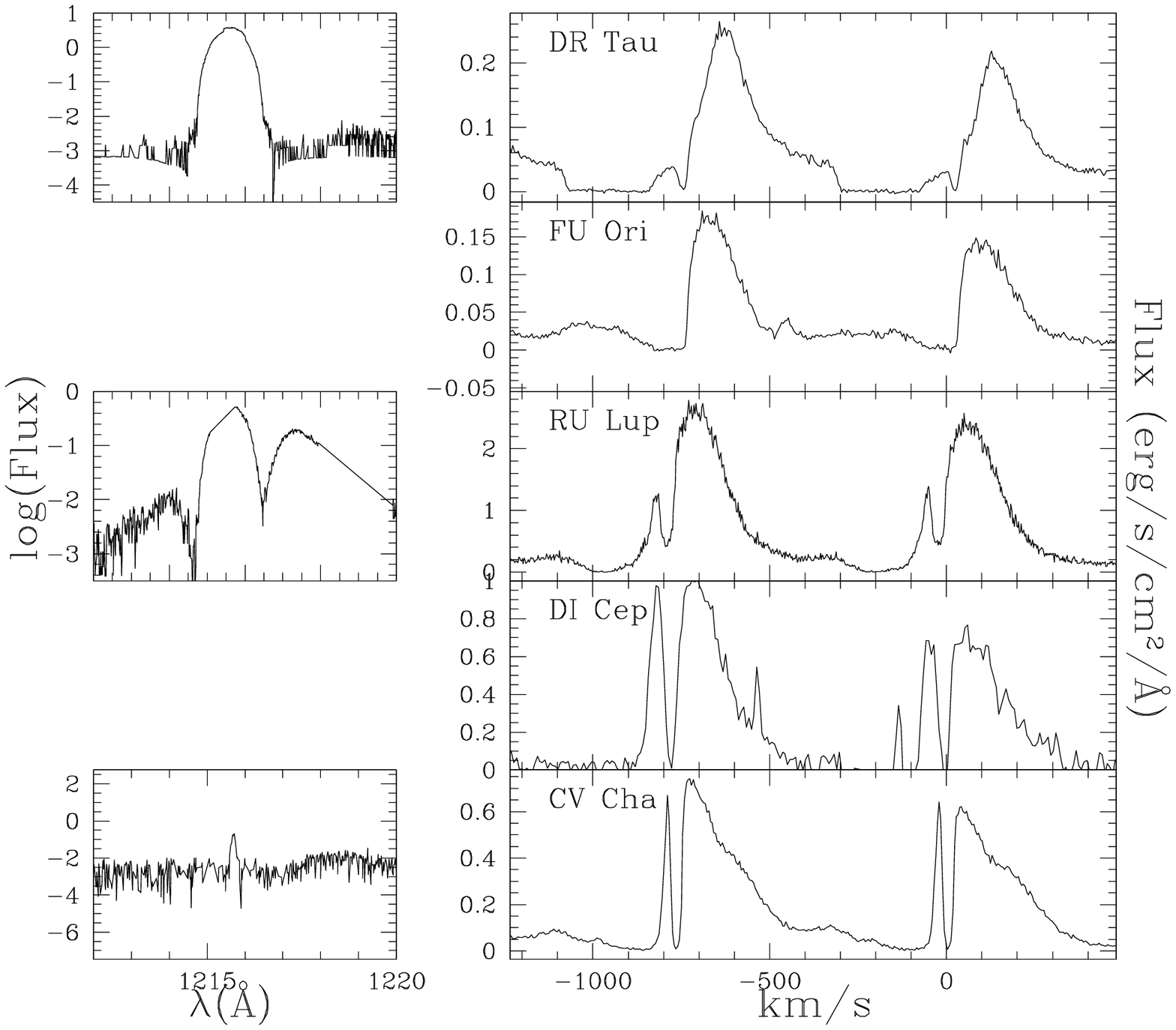} & \includegraphics[width=8cm]{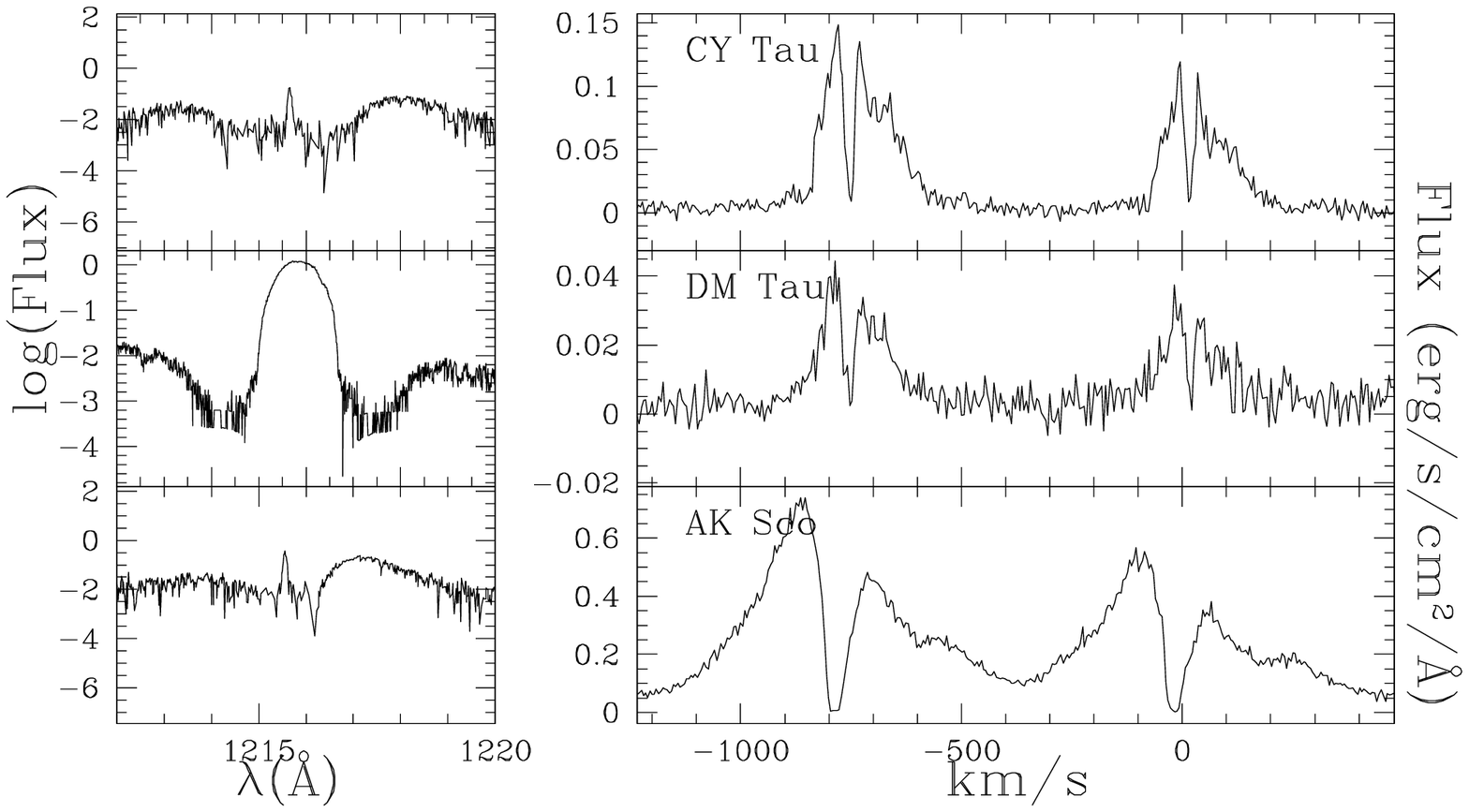} \\
\end{tabular}
\contcaption{}
\end{figure*}
%%%%%%%%%%%%%%%%%%%%%%%

\section{General properties of the sample}
\label{generalproperties}
The sample covers a broad range of stellar and disc properties as summarized in Table~\ref{tabbiblio}, such as
Spectral Type (S.T.), distance (d), inclination (i), extinction value (A$_V$), accretion rate ($\dot{M}$) and stellar rotation ($v{\rm sin}(i)$).
%The location of the stars in the J-H, H-K colour-colour diagram is represented in Fig.~\ref{colorcolor}. The infrared colours were retrieved from the 2MASS catalogue for most of the sources
%\citep[see][compilation for the bright sources]{aig1997}. The location of the main sequence is defined from a sample of nearby bright X-ray sources \citep{lopezsantiago2007}, with spectral types G0 to M5. The TTSs are located in a range that agrees well with the derived by \citet{meyer1997}; note that these authors interpreted it as a track marking the evolution of the accretion discs towards the main sequence.  

%It is worth noticing the dispersion in J-H for H-K $\sim 0.58$. This cannot be caused by extinction (note the reddening direction
%in the figure). In fact, it is caused by the broad range of spectral types (G to M) and photospheric effective temperatures covered by the sample. As
%shown in Fig.~\ref{excesocolor}, where infrared colours are corrected from the photospheric contribution, the dispersion decreases significantly. To apply this correction, we used the main sequence infrared colours from \citet{ducati2001}.
%%%%%%%%%%%%%%%%%%%%%%
\begin{table*}
\caption{Properties of the stars analysed in this work.}
\label{tabbiblio}
\centering
\begin{tabular}{lllllllllll}
\hline
Star     & S.T.&  d  &    $i$         &$A_V$   &$\dot{M}(M_{\sun} yr^{-1})$ & $v{\rm sin}(i)$ & $\log (T_{eff}/K)$   & $\log (L/L_{\sun})$   & $M$  & Age \\
  &   & (pc)   & (deg)        & (mag)    & ($\times 10^{-8}$)	& (km/s)& & & $(M_{\sun})$ & (Myr) \\
\hline
AA	Tau	&	K7	$^{(24)}$	&	140	&	75	$^{(18)}$	&	1.9	$^{(24)}$	&	1.50	$^{(24)}$	&	11	$^{(1)}$	&	3.58	$^{(32)}$	&	-0.19	$^{(32)}$	&	0.4		&	0.6		\\
AK	Sco	&	F5	$^{(12)}$	&	145	&	65-70	$^{(12)}$	&	0.5	$^{(12)}$	&	-	&	18.5	$^{(12)}$	&	3.81	$^{(3)}$	&	0.82	$^{(31)}$	&	1.5	$^{(a)}$	&	20.0		\\
BP	Tau	&	K7	$^{(24)}$	&	140	&	30	$^{(18)}$	&	1.1	$^{(24)}$	&	2.90	$^{(24)}$	&	10	$^{(1)}$	&	3.61	$^{(19)}$	&	-0.19	$^{(19)}$	&	0.5		&	1.1		\\
CS	Cha	&	K6	$^{(24)}$	&	160	&	60	$^{(18)}$	&	0.3	$^{(24)}$	&	0.53	$^{(24)}$	&	21	$^{(3)}$	&	3.64	$^{(3)}$	&	0.43	$^{(3)}$	&	0.6		&	0.3		\\
CV	Cha	&	G9	$^{(24)}$	&	160	&	35	$^{(10)}$	&	1.5	$^{(24)}$	&	5.90	$^{(24)}$	&	32	$^{(3)}$	&	3.74	$^{(3)}$	&	0.90	$^{(3)}$	&	3.0		&	1.5	$^{(a)}$	\\
CY	Tau	&	M2	$^{(4)}$	&	140	&	47	$^{(25)}$	&	0.03	$^{(4)}$	&	0.14	$^{(4)}$	&	10.6	$^{(15)}$	&	3.57	$^{(19)}$	&	-0.40	$^{(19)}$	&	0.4		&	1.1		\\
DE	Tau	&	M2	$^{(24)}$	&	140	&	35	$^{(18)}$	&	0.9	$^{(24)}$	&	2.80	$^{(24)}$	&	10	$^{(1)}$	&	3.55	$^{(3)}$	&	-0.04	$^{(3)}$	&	0.3		&	0.1		\\
DF	Tau	&	M1	$^{(4)}$	&	140	&	80	$^{(6)}$	&	0.15	$^{(4)}$	&	1.00	$^{(4)}$	&	16.1	$^{(27)}$	&	3.57	$^{(19)}$	&	-0.33	$^{(19)}$	&	0.4		&	0.8		\\
DG	Tau	&	K6	$^{(4)}$	&	140	&	90	$^{(7)}$	&	1.41	$^{(4)}$	&	4.60	$^{(4)}$	&	20	$^{(3)}$	&	3.62	$^{(30)}$	&	-0.55	$^{(30)}$	&	0.8		&	9.0		\\
DI	Cep	&	G8IV	$^{(13)}$	&	300	&	-		&	0.24	$^{(13)}$	&	$\ga$0.6	$^{(13)}$	&	23.5	$^{(37)}$	&	3.74	$^{(3)}$	&	0.71	$^{(3)}$	&	2.0		&	3.0		\\
DK	Tau	&	K7	$^{(24)}$	&	140	&	50	$^{(18)}$	&	1.3	$^{(24)}$	&	3.40	$^{(24)}$	&	11.5	$^{(1)}$	&	3.61	$^{(19)}$	&	-0.05	$^{(19)}$	&	0.5		&	0.6		\\
DM	Tau	&	M1	$^{(24)}$	&	140	&	35	$^{(18)}$	&	0.7	$^{(24)}$	&	0.29	$^{(24)}$	&	4	$^{(15)}$	&	3.57	$^{(19)}$	&	-0.80	$^{(19)}$	&	0.6		&	7.0		\\
DN	Tau	&	M0	$^{(24)}$	&	140	&	28	$^{(18)}$	&	0.9	$^{(24)}$	&	1.00	$^{(24)}$	&	12.3	$^{(15)}$	&	3.59	$^{(19)}$	&	-0.10	$^{(19)}$	&	0.4		&	0.5		\\
DR	Tau	&	K5	$^{(24)}$	&	140	&	72	$^{(18)}$	&	1.4	$^{(24)}$	&	5.20	$^{(24)}$	&	10.0	$^{(3)}$	&	3.61	$^{(19)}$	&	0.29	$^{(19)}$	&	0.4		&	0.2		\\
DS	Tau	&	K5	$^{(4)}$	&	140	&	90	$^{(25)}$	&	0.9	$^{(4)}$	&	1.20	$^{(4)}$	&	10.0	$^{(1)}$	&	3.69	$^{(3)}$	&	-0.22	$^{(3)}$	&	1.1		&	12.0		\\
FM	Tau	&	M0	$^{(24)}$	&	140	&	-		&	0.7	$^{(24)}$	&	0.12	$^{(24)}$	&	-		&	3.50	$^{(32)}$	&	-0.65	$^{(32)}$	&	0.1		&	0.3		\\
FU	Ori	&	G0	$^{(38)}$	&	450	&	40	$^{(8)}$	&	-		&	-		&	-		&	-		&	-		&	-		&	-		\\
GM	Aur	&	K7	$^{(24)}$	&	140	&	55	$^{(18)}$	&	0.6	$^{(24)}$	&	0.96	$^{(24)}$	&	12.4	$^{(27)}$	&	3.68	$^{(19)}$	&	0.09	$^{(19)}$	&	1.0		&	2.5		\\
GW	Ori	&	G0	$^{(26)}$	&	450	&			&	1.3	$^{(26)}$	&	27.00	$^{(26)}$	&	40	$^{(26)}$	&	3.75	$^{(3)}$	&	1.82	$^{(3)}$	&	3.0		&	1.0	$^{(a)}$	\\
HBC	388$^{(W)}$&	K1	$^{(4)}$	&	140	&	45	$^{(6)}$	&	0	$^{(4)}$	&	0.40	$^{(4)}$	&	19.5	$^{(16)}$	&	3.71	$^{(2)}$	&	0.15	$^{(2)}$	&	1.4		&	5.0		\\
HBC	427$^{(W)}$	&	K5	$^{(17)}$	&	140	&	67	$^{(20)}$	&	0	$^{(17)}$	&	-		&	-		&	3.64	$^{(20)}$	&	-0.12	$^{(20)}$	&	0.7		&	1.9		\\
HN	Tau	&	K5	$^{(24)}$	&	140	&	45	$^{(18)}$	&	1.1	$^{(24)}$	&	1.40	$^{(24)}$	&	52.8	$^{(27)}$	&	3.60	$^{(32)}$	&	-0.56	$^{(32)}$	&	0.7		&	6.0		\\
IP	Tau	&	M0	$^{(24)}$	&	140	&	60	$^{(18)}$	&	1.7	$^{(24)}$	&	0.72	$^{(24)}$	&	12.3	$^{(15)}$	&	3.58	$^{(19)}$	&	-0.36	$^{(4)}$	&	0.5		&	1.5		\\
LkCa	4$^{(W)}$	&	K7	$^{(4)}$	&	140	&	-		&	1.21	$^{(4)}$	&	0.19	$^{(4)}$	&	30	$^{(15)}$	&	3.61	$^{(19)}$	&	-0.13	$^{(19)}$	&	0.5		&	0.9		\\
LkCa	19$^{(W)}$	&	K0	$^{(4)}$	&	140	&	-		&	0.74	$^{(4)}$	&	0.01	$^{(4)}$	&	21	$^{(3)}$	&	3.72	$^{(19)}$	&	0.19	$^{(19)}$	&	1.4		&	9.0		\\
PDS	66	&	K1	$^{(24)}$	&	86	&	30	$^{(22)}$	&	0.2	$^{(24)}$	&	0.01	$^{(24)}$	&	14	$^{(21)}$	&	3.70	$^{(35)}$	&	0.00	$^{(35)}$	&	1.2		&	7.0		\\
RECX	1$^{(W)}$	&	K4	$^{(25)}$	&	97	&	-		&	0	$^{(25)}$	&	-		&	22	$^{(21)}$	&	3.63	$^{(21)}$	&	0.00	$^{(24)}$	&	0.6		&	0.8		\\
RECX	15	&	M3	$^{(24)}$	&	97	&	60	$^{(18)}$	&	0	$^{(24)}$	&	0.08	$^{(24)}$	&	15.9	$^{(28)}$	&	3.53	$^{(28)}$	&	-1.07	$^{(28)}$	&	0.3		&	5.0		\\
RECX	11	&	K5	$^{(24)}$	&	97	&	70	$^{(18)}$	&	0	$^{(24)}$	&	0.02	$^{(24)}$	&	16.4	$^{(29)}$	&	3.65	$^{(34)}$	&	-0.22	$^{(24)}$	&	0.9		&	4.0		\\
RU	Lup	&	K7	$^{(17)}$	&	140	&	24	$^{(11)}$	&	0.1	$^{(17)}$	&	-		&	9	$^{(11)}$	&	3.61	$^{(30)}$	&	-0.38	$^{(30)}$	&	0.6		&	2.8		\\
RW	Aur	&	K3	$^{(24)}$	&	140	&	40	$^{(6)}$	&	0.5	$^{(24)}$	&	2.00	$^{(24)}$	&	15	$^{(3)}$	&	3.66	$^{(19)}$	&	0.24	$^{(19)}$	&	0.8		&	0.8		\\
RY	Tau	&	G1	$^{(9)}$	&	140	&	86	$^{(5)}$	&	2.2	$^{(9)}$	&	6.80	$^{(9)}$	&	48.7	$^{(1)}$	&	3.71	$^{(19)}$	&	0.82	$^{(19)}$	&	1.9		&	2.4	$^{(a)}$	\\
S	CrA	&	K6	$^{(3)}$	&	130	&	-		&	0.5	$^{(3)}$	&	-		&	-		&	3.63	$^{(3)}$	&	0.11	$^{(3)}$	&	0.6		&	0.5		\\
SU	Aur	&	G1	$^{(9)}$	&	140	&	86	$^{(5)}$	&	0.9	$^{(9)}$	&	4.90	$^{(9)}$	&	59	$^{(15)}$	&	3.77	$^{(19)}$	&	0.97	$^{(19)}$	&	2.0		&	6.3	$^{(a)}$	\\
SZ	102	&	K0	$^{(25)}$	&	200	&	10	$^{(18)}$	&	0.32	$^{(25)}$	&	0.79	$^{(25)}$	&	-		&	3.72	$^{(23)}$	&	-1.94	$^{(23)}$	&	-		&	-		\\
T	Tau	&	K0	$^{(4)}$	&	140	&	20	$^{(6)}$	&	1.46	$^{(4)}$	&	3.20	$^{(4)}$	&	20.1	$^{(1)}$	&	3.72	$^{(3)}$	&	0.90	$^{(3)}$	&	2.0		&	1.8	$^{(a)}$	\\
TW	Hya	&	K7	$^{(24)}$	&	56	&	7	$^{(7)}$	&	0	$^{(24)}$	&	0.18	$^{(24)}$	&	5.8	$^{(7)}$	&	3.61	$^{(30)}$	&	-0.77	$^{(30)}$	&	0.8		&	20.0		\\
TWA	7$^{(W)}$	&	M1	$^{(17)}$	&	27	&	28	$^{(14)}$	&	0	$^{(17)}$	&	-		&	4	$^{(21)}$	&	3.52	$^{(14)}$	&	-0.49	$^{(33)}$	&	0.2		&	0.3		\\
TWA	3A	&	M3	$^{(24)}$	&	50	&	-		&	0	$^{(24)}$	&	0.01	$^{(24)}$	&	12	$^{(21)}$	&	3.53	$^{(21)}$	&	-1.10	$^{(17)}$	&	0.3		&	5.0		\\
TWA 13A	$^{(W)}$	&	M1	$^{(17)}$	&	53	&	-		&	0	$^{(17)}$	&	-		&	12	$^{(21)}$	&	3.56	$^{(36)}$	&	-0.79	 $^{(33)}$	&	0.5		&	6.0		\\
UX	Tau	&	K5	$^{(4)}$	&	140	&	35	$^{(18)}$	&	0.26	$^{(4)}$	&	1.10	$^{(25)}$	&	25.4	$^{(39)}$	&	3.64	$^{(19)}$	&	0.11	$^{(19)}$	&	0.7		&	0.7		\\
V819	Tau$^{(W)}$	&	K7	$^{(4)}$	&	140	&	-		&	1.64	$^{(4)}$	&	0.14	$^{(4)}$	&	9.1	$^{(15)}$	&	3.60	$^{(32)}$	&	-0.13	$^{(32)}$	&	0.5		&	0.7		\\
V836	Tau	&	K7	$^{(24)}$	&	140	&	65	$^{(18)}$	&	1.5	$^{(24)}$	&	0.11	$^{(24)}$	&	13.4	$^{(15)}$	&	3.61	$^{(30)}$	&	-0.49	$^{(30)}$	&	0.7		&	5.0		\\

\hline
\end{tabular}
\begin{flushleft}
(a) Values taken from other authors because they could not be measured from \citet{dantona1997} tracks.\\
(W) WTTSs of the sample according to the references indicated in this table.\\

(1) \citet{hartmann1986}; (2) \citet{kundurthy2006}; (3) \citet{johnskrull2000};
(4) \citet{white2001}; (5) \citet{muzerolle2003};
(6) \citet{ardila2002}; (7) \citet{herczeg2006}; 
(8) \citet{hartmann2004}; 
(9) \citet{salyk2013}; (10) \citet{hussain2009}; 
(11) \citet{stempels2007}; (12) \citet{aig2009b};
(13) \citet{aig1996}; (14) \citet{yang2008};
(15) \citet{nguyen2009}; (16) \citet{sartoretti1998}; 
(17) \citet{yang2012}; (18) \citet{france2012}; (19) \citet{bertout2007}; 
(20) \citet{steffen2001};  (21) \citet{dasilva2009};
(22) \citet{sacco2012}; (23) \citet{hughes1994}; (24) \citet{ingleby2013};
(25) \citet{ardila2013}; (26) \citet{calvet2004}; 
(27) \citet{clarke2000}; (28) \citet{woitke2011}; 
(29) \citet{jayawardhana2006}; 
(30) \citet{herczeg2008}; (31) \citet{manoj2006}; 
(32) \citet{hartigan1995};  (33) \citet{ingleby2011}; 
(34) \citet{lawson2001};
(35) \citet{mamajek2002}; (36) \citet{sterzik1999}; (37) \citet{azevedo2006}; 
(38) \citet{petrov2008}; (39) \citet{preibisch1997}.
\end{flushleft}
\end{table*}
%%%%%%%%%%%%%%%%
 
%%%%%%%%%%%%%%%%%% 

%\begin{figure}
%\centering
%\includegraphics[width=8cm]{fig2.eps}
%\caption{Colour-colour diagram for TTSs of the sample. The main sequence stars location \citep{lopezsantiago2007} are represented by diamonds, reddening vector
%and the \citet{meyer1997} fit are plotted for reference. The triangles are WTTSs.\label{colorcolor}} 
%\end{figure}
%%%%%%%%%%%%%%%%%
%\begin{figure}
%\centering
%\includegraphics[width=8cm]{fig3.eps}
%\caption{Colour excess with respect to the main sequence stars for TTSs in the sample. WTTSs are represented by triangles. \label{excesocolor}}
%\end{figure}
%%%%%%%%%%%%%%%%

\subsection{Age and mass}

Age and mass determinations for TTSs are uncertain. Published measurements of TTSs luminosities and effective temperatures (see Table~\ref{tabbiblio}) were used to compute the masses and ages provided in the last two columns of Table~\ref{tabbiblio}. 

\begin{figure}
\centering
\includegraphics[width=8cm]{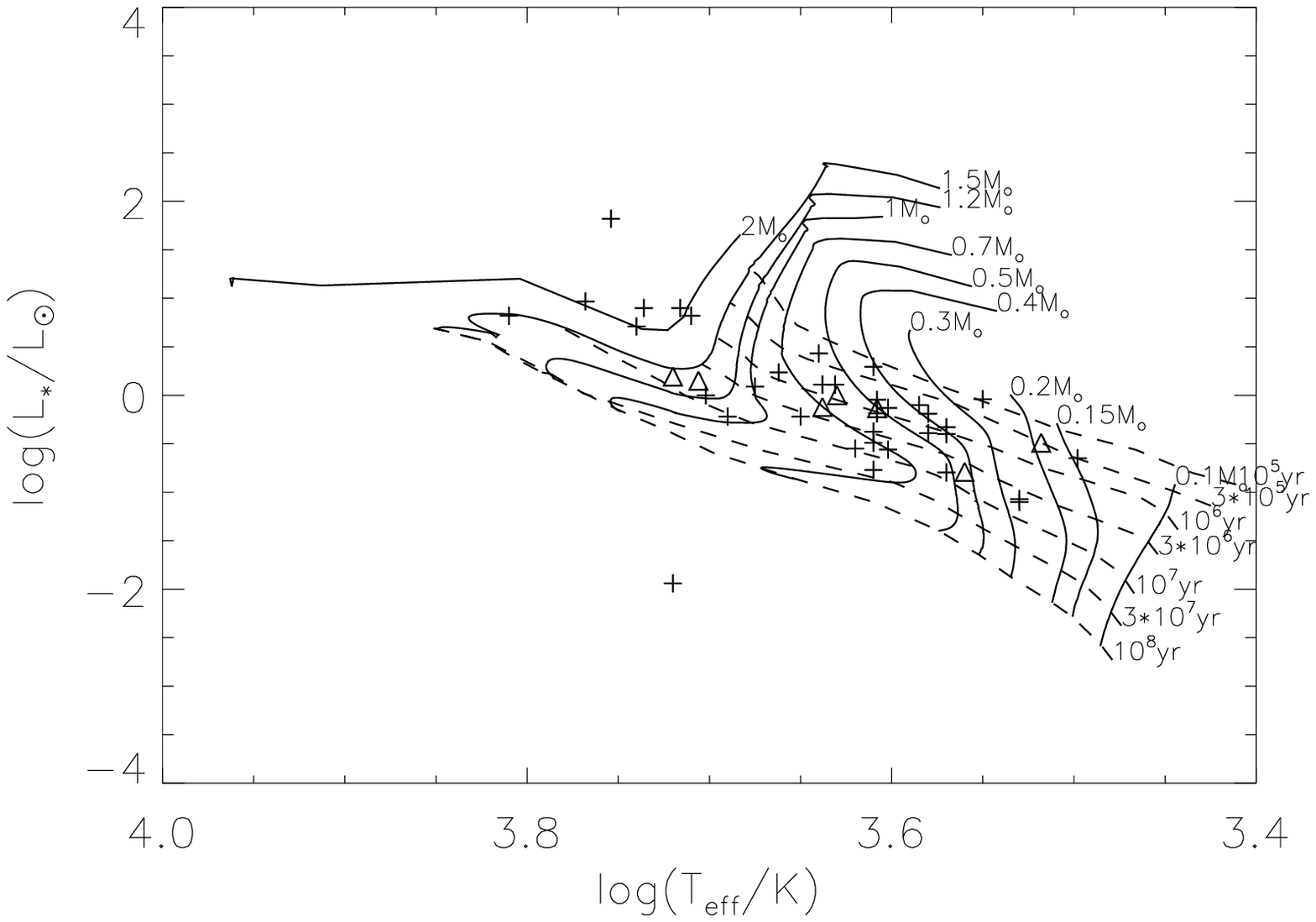}\\
\includegraphics[width=8cm]{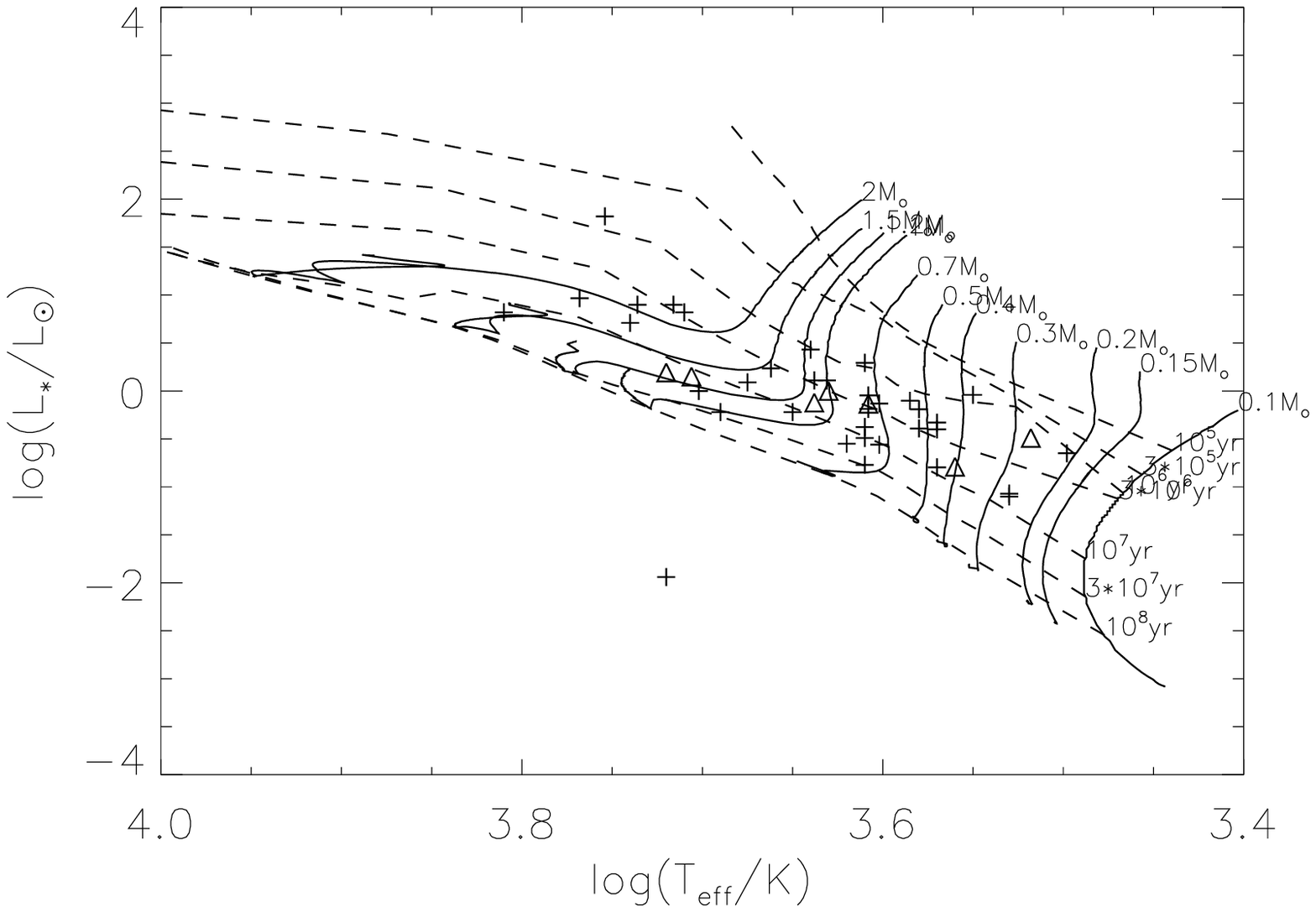}
\caption{Location of the TTSs on the \citet{dantona1997} (top) and \citet{siess2000} (bottom) PMS evolutionary tracks. Triangles represent WTTSs. \label{isocronas}}
\end{figure}

First, we used the \citet{dantona1997} evolutionary tracks (see Fig.~\ref{isocronas}, top panel). This model 
introduces a Kolmogorov based turbulence cascade \citep{canuto1991} in the parametrisation of the
internal stellar heat transport. According to these tracks, our sample covers the age range from 1 to 10 Myr
and a broad range of masses, from 0.2 to 2 M$_{\odot}$. 

Secondly, we used the \citet{siess2000} evolutionary tracks (see Fig.~\ref{isocronas}, bottom panel). These evolutionary
tracks do not include a complex treatment of transport but take into account the effect of fresh deuterium accretion in PMS evolution. According to these tracks, the range of masses of the stars in our sample is reduced with respect to the D'Antona's tracks but the spread in age is increased notably: from 3 to 30~Myr. 

\begin{figure}
\centering
\includegraphics[width=8cm]{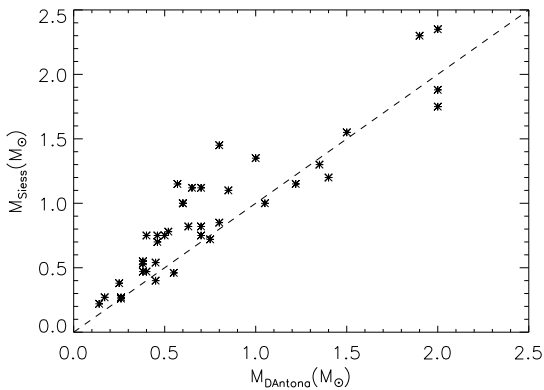}\\
\caption{Comparison of the mass estimates obtained using \citet{dantona1997} and \citet{siess2000}
PMS evolutionary tracks. The dashed line represents $M_{D'Antona}=M_{Siess}$. \label{masasyedades}}
\end{figure}

Finally, in Fig.~\ref{masasyedades}, masses are compared for both sets of estimates. Note that the mass calculations are rather robust, i.e., both sets of evolutionary tracks provide similar results. 
We also compared the ages for both PMS evolutionary tracks. We found a
large discrepancy in the age estimates.
These masses and ages values are provided in the last two columns of Table~\ref{tabbiblio}. 
%%%%%%%%%%%%%%%%
%%%%%%%%%%%%%%%%
%%%%%%%%%%%%%%%%

\subsection{Binaries}

There are several binaries and multiples in the sample: RY Tau \citep{bertout1999}, AK Sco \citep{andersen1989}, DF~Tau \citep{bertout1988,bouvier1993,unruh1998}, GW~Ori \citep{mathieu1991}, CV~Cha \citep{bertout1999,hussain2009}, RW~Aur \citep{bertout1999,white2001}, S~CrA \citep{walter2005}, DK~Tau, HN~tau \citep{correia2006}, UX~Tau, V819~Tau \citep{nguyen2012}, FU~Ori \citep{wang2004}, RECX~1, HBC~427, CS~Cha \citep{ardila2013} and T~Tau \citep{furlan2006,herbst1996}. 

Four of them, namely RY~Tau, AK~Sco, DF~Tau and GW~Ori are close binaries with semi-major axes 3.17, 0.14, 12.6 and 1~AU, respectively. 
Henceforth the contribution from the components is unresolved in the \textit{IUE} and \textit{HST}/GHRS profiles. 

The distances between the components in the  CV~Cha, S~CrA, RW~Aur, DK~Tau, HN~Tau, UX~Tau, FU~Ori, T~Tau, RECX~1, HBC~427 and V819~Tau are $11.4$, $1.3$, $1.4$, $2.3$, $3.1$, $5.9$, $0.5$, $0.7$, $0.18$, $0.03$ and $10.5$ arcsec, respectively. RW~Aur, UX~Tau \citep{nguyen2012}, RECX~1, T~Tau \citep{ardila2013}, and GW~Ori \citep{berger2011} are multiple systems.

\subsection{Optical jets sources}

RY~Tau \citep{stonge2008}, RW~Aur \citep{white2001,coffey2008}, DG~Tau \citep{white2001,coffey2008}, and T~Tau \citep{furlan2006,herbst1996} are sources of resolved jets. The inclinations of the jets of DG Tau and RW Aur with respect to the plane of the sky have been estimated to be $52^\circ$ and $44^\circ$, respectively. DG~Tau jet is well collimated with knots and bow shocks out to at least $11$~arcsec, with velocities of several 100 Km~s$^{-1}$ \citep{gudel2008}. The Mg~II emissions from RW~Aur, HN~Tau, DP~Tau and CW~Tau jets have been measured \citep{coffey2012}. The Mg~II lines are roughly 1-2 orders of magnitude stronger than the optical forbidden lines and, in general, the approaching jet is brighter than the receding jet, as otherwise expected by the impact of circumstellar extinction and the absorption by the intervening warm environment. The (unresolved) jet contribution to the Mg~II profile is shown in the high velocity  edge of the profile \citep[see][]{coffey2008}.

\subsection{Magnetic fields and spots}

Strong magnetic fields have been directly detected in
CTTSs. Magnetic fields of kilo-Gauss~(kG) have been measured from Zeeman
broadening measurements only for few  sources: AA~Tau (2.78~kG), BP~Tau (2.17~kG), CY~Tau (1.16~kG), DE~Tau (1.12~kG), DF~Tau (2.90~kG),
DG~Tau (2.55~kG), DK~Tau (2.64~kG), DN~Tau (2.00~kG), GM~Aur (2.22~kG), T~Tau (2.37~kG) and TW~Hya (2.61~kG) \citep{johnskrull2007}. Indications of 
strong magnetic activity or magnetic channelled accretion have been found in some other sources.

The study of the Zeeman broadening analysis and measurement of the circular
polarization signal allows to derive the field topology itself. Magnetic surface maps
have been published for several accreting TTSs derived from the
technique of Zeeman-Doppler imaging \citep[see, for instance,][]{donati2007,donati2008,donati2012,hussain2009,gregory2012}.
We use for CV~Cha the magnetic field derived with this technique: $>0.02$~kG.
%spots

Hot spots on the stellar surface are produced by accretion shocks and they have been detected in BP~Tau \citep{aig1997,johnskrull2004,donati2008}, CY~Tau \citep{bouvier1995}, DF~Tau \citep{bertout1988,bouvier1993,unruh1998} and DI~Cep \citep{aig1996}.

\section{Measurements and data analysis}
\label{measurements}
Fig.~\ref{profiles} shows a trend of increasing Mg~II strength and profile broadening from WTTSs to classical TTSs. 
There is not a clear cut separation between the two groups. Rather, it seems there is a sequence
associated with the line emitting volume and the strength of the wind. The correlation between broadening and strength has been already noticed for other spectral tracers, such as the C~IV or the N~V lines, most recently 
by \citet{ardila2013,aig2013b}. This sequence is also associated with the weakening of the 
H$_2$ molecular emission and the evaporation of the gas in the circumstellar disc. 

Let us follow the trend outlined in Fig.~\ref{profiles}. The Mg~II profiles of the WTTSs are rather narrow (typical widths at the base of the line are $\sim 160.5$ km~s$^{-1}$) with a circumstellar absorption feature over the TTS line emission. It is noticeable that the feature is at rest with respect to the star's emission only in HBC~427 and V819~Tau. In LkCa~19 and LkCa~4, it is slightly red-wards shifted and in the rest of the sources blue-wards shifted. Given the strength of the feature and the location of our sample stars (at distances
smaller than 200~pc from the Sun for the majority of the sources), the feature is expected to be produced by warm absorbing circumstellar material. These slight shifts suggest a mainly outflow motion in the circumstellar environment along the line of sight. 

The comparison with the Ly-$\alpha$ profiles also shows the uncertainties of a morphologically based classification in terms of the Mg~II profile. The Ly-$\alpha$ profiles of the WTTSs (LkCa~4, LkCa~19 and HBC~427) are narrow enough to be fully covered by the geocoronal Ly-$\alpha$ emission; no high velocity wings are detected (see Fig.~\ref{profiles}). However, TTSs with apparently similar Mg~II profiles, such as FM~Tau, TWA~7, TWA~13A, RECX~1, RECX~11 display broad wings in Ly-$\alpha$.

Mg~II profiles in CTTSs can be described as a broad emission with significant wind absorption in the blue wing (in addition to the narrow circumstellar feature). The high S/N of the \textit{HST} observations permit to follow the velocity law in the wind and its geometry. The strength of the wind varies significantly from source to source and, in some objects like GM~Aur, two broad absorption components (winds?) are observed. There are three peculiar profiles in the sample: BP~Tau, RW~Aur and AK~Sco. There is not significant wind absorption in BP~Tau. RW~Aur profile is extremely broad; this fact is even noticeable in tracers like the C~III] and Si~III]  semi forbidden transitions and drove \citet{aig2003} to hypothesize the existence
of an ionized plasma torus around this star. AK~Sco is the only star displaying a broad red-wards shifted absorption but this is caused by the complex circumstellar gas dynamics in this close binary system \citep[see the numerical simulations in][]{aig2013c}.
In the following, we describe the procedures followed to quantify the evolution of the profiles and the characteristics of the outflows.

\subsection{ Mg~II flux measurements and flux-flux correlations}
\label{fluxes}
The flux radiated in the Mg~II lines is calculated as
$F_{a,b} = (\sum_{i=0}^{i=n _{a,b}} F_{a,b}(\lambda _i) - n_{a,b} < C >)* \delta \lambda, $
\noindent
being $\lambda _{0,a}$ and $\lambda _{0,b}$ the short wavelength edges of the 2796~\AA\ and 2804~\AA\ Mg~II lines (subscripts 'a' and 'b', respectively). $\delta\lambda$ is the pixel size, $\lambda _i = \lambda _0 + i*\delta \lambda$ and $n_{a,b}$ the number of pixels in each profile.  The average continuum level $ < C > $ was determined in one nearby, featureless, window and the dispersion about this average, $\sigma$, is used to compute the flux errors as $d F_{a,b} = \sigma n_{a,b} $ (see below for the determination of the profile edge). 
These flux errors are represented in figures with error bars.
For the measured velocities we took an error of 1~\AA\ for all stars.
Since the stellar bolometric, $F_{bol}$,
is $F_{bol}=(\sigma T^4 4 {\rm \pi} R_*^2)/4 {\rm \pi} d^2$,
the rate $F_l/F_{bol}$, with l the corresponding line, provides
a measure of the line emissivity corrected from stellar radii and
surface temperature. In this manner, the normalised
fluxes are corrected from scaling effects associated with
the broad range of mass, luminosity and stellar radius
covered by the TTSs sample studied in this work.

\begin{figure}
\centering
\includegraphics[width=9cm]{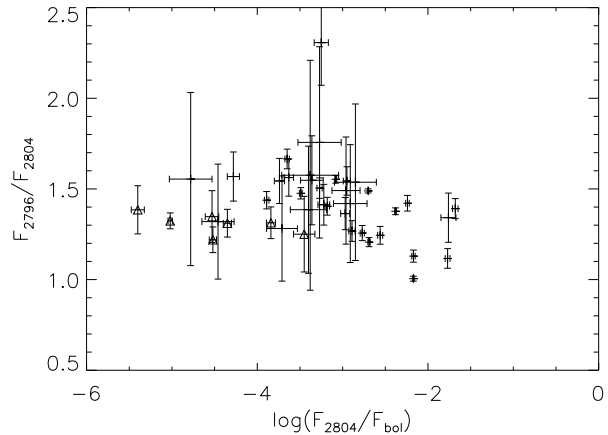}
\caption{The 2796/2804 flux ratio, as a function of the flux of the Mg~II($\lambda $ 2804~\AA) line. WTTSs are represented by triangles.\label{saturacion} }
\end{figure}

As shown in Fig.~\ref{saturacion}, the ratio $F_{2796}/F_{2804} < 2$ in most sources; the line is optically thick\footnote{
The Mg~II[uv1] multiplet corresponds to transitions 2P$_0^{1/2,3/2} \rightarrow $2S$^{1/2}$ with transition
probabilities for J=3/2,1/2$ \rightarrow 1/2$  of $2.60\times 10^{8}\ s^{-1}$ and 
$2.57\times 10^{8}\ s^{-1}$, respectively.} both in  WTTSs to CTTSs. 
The average value is $1.4 \pm 0.2$. Thus, though the 
2796~\AA\ line is saturated in all sources, the 2804~\AA\ line may be not it in many observations. For this reason, 
all tests for flux scaling with other parameters are carried out  using the 2804~\AA\ line.
The 2795~\AA\ line will only be used in this article to determine some kinematical properties of the outflow
that are not affected by the saturation.  Note that this also means that scaling based on Mg~II fluxes determined from 
low dispersion data should be treated with care.

All line fluxes are provided in Table~\ref{tabfluxes}. Fluxes were extinction corrected using \citet{valencic2004} extinction law (assuming $R_V=\frac{A_V}{E(B-V)}=3.1)$ and the $A_V$ values in Table~\ref{tabbiblio}.  
Note that there are large differences in the extinctions quoted by different authors \citep[compare for instance][]{ingleby2011,yang2012}.
We estimate that extinctions are uncertain by  $\leq 0.5$ mag which corresponds to a flux 
uncertainty by a factor $\leq 1.5$.  

\begin{figure}
\centering
\includegraphics[width=9cm]{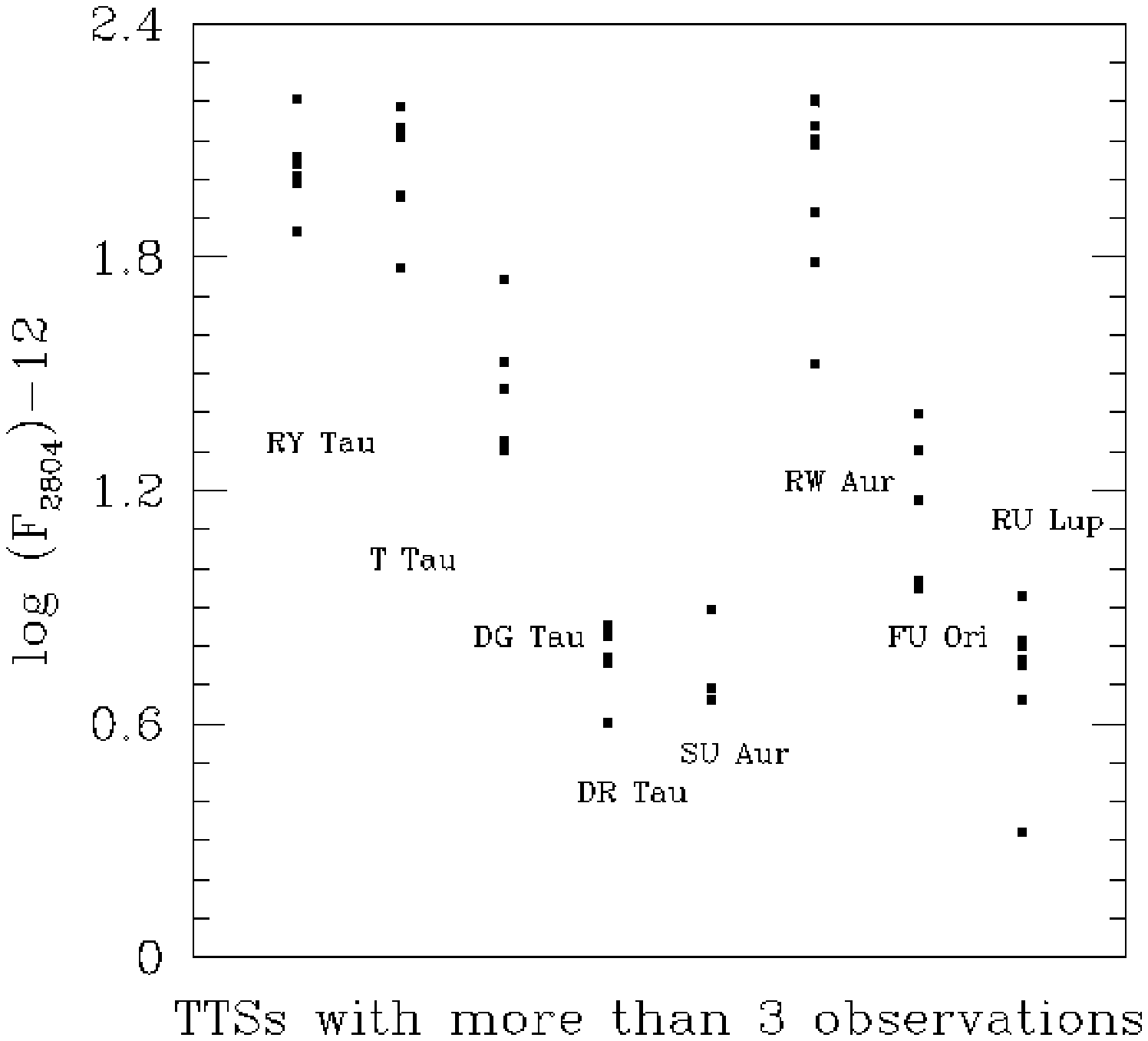}
\caption{Mg~II flux variability for the TTSs with multiple (more than three) good observations. \label{observations}}
\end{figure}

From Table~\ref{tabfluxes}, it can be readily inferred variations in the line flux by factors of $\sim 5$ over the years. 
This is observed in all the TTSs for which there are several observations available. In Fig.~\ref{observations} we have represented those 
being observed more than three times. Note that the typical observing times are long (see Table~\ref{logofobservations}), 
so flares and eruptive events can be excluded in most cases.
%%%%%%%%%%
%%%%%%%%%%
%%%%%%%%%%%%%%%%%%%%%%
\begin{table*}
\caption{Mg~II flux measurements corrected from extinction and velocity measurements of the analysed profiles.}
\label{tabfluxes}
\centering
\begin{tabular}{lllllllll}
\hline
    &   &  \multicolumn{3}{c}{Flux Measurements} & &\multicolumn{3}{c}{Velocity Measurements}\\
 Star & Date & $F_{2796}$ & $F_{2804,b}$ & $F_{2804,r}$ & & $V_{term}$ & DACs  & Comments  \\
\cline{3-5} \cline{7-9} \\
 & (yy-mm-dd)  &  \multicolumn{3}{c}{($10^{-12} erg\ s^{-1}\ cm^{-2}$)} & &\multicolumn{2}{c}{(km/s)}& \\
\hline
AA	Tau	&	07-11-01	&	$	1.93	\pm	0.46	$	&	$	0.63	\pm	0.28	$	&	$	0.74	\pm	0.26	$	&	&	-50.0		&		&	narrow	\\	\hline
AK	Sco	&	86-08-06	&	$	3.04	\pm	0.39	$	&	$	1.35	\pm	0.17	$	&	$	0.76	\pm	0.22	$	&	&			&	129.8	&	no absorption	\\	
		&	88-04-02	&	$	3.20	\pm	0.59	$	&	$	0.92	\pm	0.20	$	&	$	0.45	\pm	0.25	$	&	&			&		&	no absorption	\\	
		&	88-04-02	&	$	2.37	\pm	0.75	$	&	$	0.79	\pm	0.33	$	&	$	0.38	\pm	0.32	$	&	&			&		&	no absorption	\\	
		&	88-04-09	&	$	4.01	\pm	0.48	$	&	$	1.16	\pm	0.18	$	&	$	1.17	\pm	0.23	$	&	&			&		&	no absorption	\\	
		&	10-08-21	&	$	3.86	\pm	0.28	$	&	$	1.60	\pm	0.11	$	&	$	0.87	\pm	0.16	$	&	&			&		&	no absorption	\\	\hline
BP	Tau	&	81-07-24	&	$	15.40	\pm	1.79	$	&	$	4.79	\pm	0.68	$	&	$	6.74	\pm	1.39	$	&	&	-213.0		&		&	variable	\\	
		&	86-10-10	&	$	12.11	\pm	2.30	$	&	$	0.08	\pm	0.06	$	&	$	8.98	\pm	1.44	$	&	&	-237.7		&		&	variable	\\	
		&	86-10-26	&	$	5.98	\pm	1.13	$	&	$	1.90	\pm	0.37	$	&	$	2.72	\pm	0.69	$	&	&	-214.0		&		&	variable	\\	
		&	93-07-30	&	$	9.01	\pm	0.40	$	&	$	3.70	\pm	0.14	$	&	$	2.69	\pm	0.13	$	&	&	-194.0		&	-95.0	&	CAD	\\	\hline
CS	Cha	&	11-06-01	&	$	0.29	\pm	0.03	$	&	$	0.05	\pm	0.01	$	&	$	0.13	\pm	0.01	$	&	&	-181.3		&	-31.5	&	narrow	\\	\hline
CV	Cha	&	79-11-11	&	$	24.16	\pm	4.85	$	&	$	4.23	\pm	1.13	$	&	$	23.48	\pm	4.82	$	&	&	-304.8		&		&	broad	\\	
		&	80-07-12	&	$	28.10	\pm	6.66	$	&	$	4.20	\pm	1.00	$	&	$	26.17	\pm	6.23	$	&	&	-336.2		&		&	broad	\\	
		&	11-04-13	&	$	25.71	\pm	0.80	$	&	$	3.95	\pm	0.27	$	&	$	17.35	\pm	0.33	$	&	&	-320.0		&		&	broad,double	\\	\hline
CY	Tau	&	00-12-06	&	$	0.19	\pm	0.02	$	&	$	0.05	\pm	0.01	$	&	$	0.07	\pm	0.01	$	&	&	-172.0		&		&		\\	
		&	00-12-06	&	$	0.18	\pm	0.02	$	&	$	0.06	\pm	0.01	$	&	$	0.08	\pm	0.02	$	&	&	-154.0		&		&	noisy	\\	\hline
DE	Tau	&	10-08-20	&	$	3.33	\pm	0.15	$	&	$	0.60	\pm	0.07	$	&	$	2.08	\pm	0.06	$	&	&	-160.0	$^{(*)}$	&	-40.5	&	sharp blue edge	\\	\hline
DF	Tau	&	93-08-08	&	$	0.53	\pm	0.06	$	&	$	0.08	\pm	0.02	$	&	$	0.30	\pm	0.03	$	&	&	-117.1		&		&	broad	\\	
		&	99-09-18	&	$	0.74	\pm	0.03	$	&	$	0.20	\pm	0.02	$	&	$	0.33	\pm	0.01	$	&	&	-130.0		&		&	broad	\\	\hline
DG	Tau	&	86-01-18	&	$	7.79	\pm	3.21	$	&	$	1.07	\pm	1.73	$	&	$	3.87	\pm	0.93	$	&	&			&		&		\\	
		&	96-02-08	&	$	14.72	\pm	0.74	$	&	$	2.42	\pm	0.36	$	&	$	8.83	\pm	0.32	$	&	&	-210.0		&	-110.0	&	broad double	\\	
		&	96-02-08	&	$	13.89	\pm	0.62	$	&	$	2.38	\pm	0.31	$	&	$	7.65	\pm	0.35	$	&	&	-210.0		&	-110.0	&	broad double	\\	
		&	01-02-20	&	$	5.76	\pm	0.32	$	&	$	0.83	\pm	0.15	$	&	$	3.70	\pm	0.15	$	&	&	-194.8		&	-87.6	&		\\	
		&	01-02-20	&	$	5.61	\pm	0.31	$	&	$	0.83	\pm	0.16	$	&	$	3.53	\pm	0.13	$	&	&	-201.3		&		&		\\	
		&	01-02-20	&	$	5.59	\pm	0.28	$	&	$	0.75	\pm	0.13	$	&	$	3.55	\pm	0.11	$	&	&	-205.6		&		&		\\	
		&	01-02-20	&	$	5.73	\pm	0.27	$	&	$	0.76	\pm	0.12	$	&	$	3.65	\pm	0.13	$	&	&	-220.0		&	-90.0	&	broad double	\\	\hline
DI	Cep	&	92-12-22	&	$	2.81	\pm	0.21	$	&	$	0.41	\pm	0.05	$	&	$	1.65	\pm	0.17	$	&	&	-349.7		&		&	broad, noisy	\\	\hline
DK	Tau	&	10-02-04	&	$	2.49	\pm	0.55	$	&	$	0.22	\pm	0.29	$	&	$	1.45	\pm	0.27	$	&	&	-206.3	$^{(*)}$	&		&		\\	\hline
DM	Tau	&	10-08-22	&	$	0.15	\pm	0.04	$	&	$	0.06	\pm	0.02	$	&	$	0.05	\pm	0.03	$	&	&			&		&	noisy	\\	\hline
DN	Tau	&	11-09-11	&	$	1.16	\pm	0.09	$	&	$	0.21	\pm	0.06	$	&	$	0.62	\pm	0.05	$	&	&	-67.1		&	-48.1	&	narrow	\\	\hline
DR	Tau	&	93-08-05	&	$	3.29	\pm	1.65	$	&	$	-1.43	\pm	0.67	$	&	$	2.70	\pm	0.41	$	&	&	-369.7		&		&	broad	\\	
		&	96-09-07	&	$	5.39	\pm	0.73	$	&	$	-1.14	\pm	0.28	$	&	$	4.69	\pm	0.25	$	&	&	-435.0		&		&	broad	\\	
		&	95-09-07	&	$	5.25	\pm	1.12	$	&	$	-1.54	\pm	0.48	$	&	$	4.87	\pm	0.42	$	&	&	-409.7		&		&	broad	\\	
		&	00-08-39	&	$	3.37	\pm	1.18	$	&	$	-1.50	\pm	0.44	$	&	$	2.68	\pm	0.37	$	&	&	-370.4		&		&	broad	\\	
		&	01-02-09	&	$	4.59	\pm	0.64	$	&	$	-1.29	\pm	0.21	$	&	$	3.34	\pm	0.18	$	&	&	-365.4		&		&	broad	\\	
		&	01-02-09	&	$	4.36	\pm	0.52	$	&	$	-1.32	\pm	0.17	$	&	$	3.24	\pm	0.14	$	&	&	-374.0		&		&	broad	\\	
		&	10-02-15	&	$	4.96	\pm	1.29	$	&	$	-1.24	\pm	0.45	$	&	$	4.25	\pm	0.42	$	&	&	-381.9		&		&	broad	\\	\hline
\end{tabular}
\end{table*}

\begin{table*}
\contcaption{}
\centering
\begin{tabular}{lllllllll}
\hline
    &   &  \multicolumn{3}{c}{Flux Measurements} & &\multicolumn{3}{c}{Velocity Measurements}\\
 Star & Date & $F_{2796}$ & $F_{2804,b}$ & $F_{2804,r}$ & & $V_{term}$ & DACs  & Comments  \\
\cline{3-5} \cline{7-9} \\
 & (yy-mm-dd)  &  \multicolumn{3}{c}{($10^{-12} erg\ s^{-1}\ cm^{-2}$)} & &\multicolumn{2}{c}{(km/s)}& \\
\hline
DS	Tau	&	00-08-24	&	$	3.25	\pm	0.28	$	&	$	0.57	\pm	0.12	$	&	$	1.62	\pm	0.13	$	&	&	-142.8	$^{(*)}$	&		&		\\	
		&	01-02-23	&	$	1.70	\pm	0.12	$	&	$	0.24	\pm	0.06	$	&	$	0.86	\pm	0.05	$	&	&	-182.0		&		&		\\	
		&	01-02-23	&	$	1.76	\pm	0.10	$	&	$	0.25	\pm	0.05	$	&	$	0.90	\pm	0.05	$	&	&	-191.3		&		&		\\	\hline
FM	Tau	&	11-09-21	&	$	0.26	\pm	0.05	$	&	$	0.10	\pm	0.02	$	&	$	0.07	\pm	0.02	$	&	&			&		&	absorbed	\\	\hline
FU	Ori	&	82-08-14	&	$	6.53	\pm	2.24	$	&	$	1.27	\pm	1.38	$	&	$	5.31	\pm	1.36	$	&	&	-173.4		&		&	broad,double	\\	
		&	83-09-05	&	$	8.00	\pm	2.35	$	&	$	1.23	\pm	1.35	$	&	$	7.86	\pm	1.52	$	&	&	-167.8		&		&	broad,double	\\	
		&	87-11-03	&	$	13.17	\pm	1.93	$	&	$	1.18	\pm	0.67	$	&	$	7.98	\pm	0.90	$	&	&	-241.3		&		&	broad,double	\\	
		&	01-02-22	&	$	4.26	\pm	0.45	$	&	$	0.06	\pm	0.23	$	&	$	3.18	\pm	0.17	$	&	&	-243.4		&	-69.4	&	broad,double	\\	
		&	01-02-22	&	$	4.24	\pm	0.33	$	&	$	0.05	\pm	0.15	$	&	$	3.24	\pm	0.14	$	&	&	-231.3		&		&	broad,double	\\	\hline
GM	Aur	&	10-08-19	&	$	1.67	\pm	0.07	$	&	$	0.34	\pm	0.04	$	&	$	0.78	\pm	0.03	$	&	&	-187.7		&	-40.8	&	narrow with tail	\\	\hline
GW	Ori	&	80-11-16	&	$	16.62	\pm	2.03	$	&	$	4.22	\pm	0.61	$	&	$	13.39	\pm	1.74	$	&	&	-280.0		&		&	narrow with tail	\\	
		&	85-10-21	&	$	37.67	\pm	1.98	$	&	$	8.36	\pm	0.72	$	&	$	21.90	\pm	1.07	$	&	&	-360.0		&		&	narrow with tail	\\	\hline
HBC	388	&	95-09-09	&	$	0.10	\pm	0.01	$	&	$	0.05	\pm	0.00	$	&	$	0.03	\pm	0.00	$	&	&			&		&	negligible	\\	\hline
HBC	427	&	11-03-30	&	$	0.05	\pm	0.01	$	&	$	0.02	\pm	0.00	$	&	$	0.02	\pm	0.00	$	&	&			&		&	negligible	\\	\hline
HN	Tau	&	10-02-10	&	$	8.91	\pm	0.70	$	&	$	2.57	\pm	0.18	$	&	$	5.48	\pm	0.27	$	&	&	-320.0		&	-130.0	&	double broad	\\	\hline
IP	Tau	&	11-03-21	&	$	1.52	\pm	0.42	$	&	$	0.47	\pm	0.24	$	&	$	0.54	\pm	0.25	$	&	&	-97.8		&		&	narrow with tail	\\	\hline
LkCa	4	&	11-03-30	&	$	0.55	\pm	0.12	$	&	$	0.23	\pm	0.05	$	&	$	0.21	\pm	0.06	$	&	&			&		&	negligible	\\	\hline
LkCa	19	&	11-03-31	&	$	0.50	\pm	0.04	$	&	$	0.20	\pm	0.02	$	&	$	0.18	\pm	0.02	$	&	&			&		&	negligible	\\	\hline
PDS	66	&	11-05-23	&	$	2.14	\pm	0.06	$	&	$	0.60	\pm	0.03	$	&	$	0.86	\pm	0.02	$	&	&	-130.0		&	-79.89;-28.16	&	double brad-narrow	\\	\hline
RECX	1	&	10-01-22	&	$	0.10	\pm	0.01	$	&	$	0.02	\pm	0.00	$	&	$	0.06	\pm	0.01	$	&	&	-70.0		&		&	narrow with tail	\\	\hline
RECX	15	&	10-02-05	&	$	0.50	\pm	0.03	$	&	$	0.13	\pm	0.01	$	&	$	0.26	\pm	0.02	$	&	&	-50.0		&	-43.0	&	narrow with tail	\\	\hline
RECX	11	&	09-12-12	&	$	0.10	\pm	0.03	$	&	$	0.03	\pm	0.01	$	&	$	0.04	\pm	0.01	$	&	&	-54.0		&		&	various comp. and long tail	\\	\hline
RU	Lup	&	81-09-11	&	$	6.19	\pm	0.60	$	&	$	-0.10	\pm	0.18	$	&	$	5.02	\pm	0.22	$	&	&	-340.0		&		&	broad P-cygni	\\	
		&	81-09-11	&	$	5.15	\pm	1.35	$	&	$	-0.10	\pm	0.69	$	&	$	5.53	\pm	0.60	$	&	&	-375.0		&		&	broad P-cygni	\\	
		&	83-04-16	&	$	4.89	\pm	1.47	$	&	$	0.24	\pm	0.20	$	&	$	4.65	\pm	0.65	$	&	&	-240.6		&		&	noisy	\\	
		&	83-04-17	&	$	5.63	\pm	0.59	$	&	$	-0.39	\pm	0.20	$	&	$	6.44	\pm	0.22	$	&	&			&		&		\\	
		&	85-07-08	&	$	6.18	\pm	0.56	$	&	$	-0.12	\pm	0.23	$	&	$	5.94	\pm	0.22	$	&	&	-380.0		&		&	broad P-cygni	\\	
		&	85-07-10	&	$	6.20	\pm	0.59	$	&	$	0.09	\pm	0.18	$	&	$	5.44	\pm	0.23	$	&	&	-400.4		&		&	broad P-cygni	\\	
		&	86-03-04	&	$	9.85	\pm	0.81	$	&	$	-0.16	\pm	0.27	$	&	$	7.72	\pm	0.28	$	&	&	-397.5		&		&	broad P-cygni	\\	
		&	88-02-19	&	$	2.32	\pm	0.20	$	&	$	0.08	\pm	0.05	$	&	$	1.77	\pm	0.08	$	&	&	-332.6		&		&	broad P-cygni	\\	
		&	92-08-24	&	$	5.50	\pm	0.24	$	&	$	0.40	\pm	0.07	$	&	$	4.48	\pm	0.12	$	&	&	-301.9		&		&	broad P-cygni	\\	\hline
RW	Aur	&	79-04-04	&	$	15.07	\pm	1.06	$	&	$	4.50	\pm	0.44	$	&	$	12.50	\pm	0.47	$	&	&	-217.7		&		&	broad	\\	
		&	79-04-09	&	$	16.50	\pm	1.87	$	&	$	6.55	\pm	0.84	$	&	$	14.28	\pm	0.91	$	&	&	-229.1		&		&	broad	\\	
		&	80-11-15	&	$	11.23	\pm	1.40	$	&	$	2.95	\pm	0.73	$	&	$	9.06	\pm	0.86	$	&	&	-214.0		&		&	broad	\\	
		&	93-08-10	&	$	17.84	\pm	0.56	$	&	$	6.24	\pm	0.18	$	&	$	10.18	\pm	0.24	$	&	&	-202.7		&		&	broad	\\	
		&	94-02-04	&	$	7.13	\pm	1.57	$	&	$	2.15	\pm	0.48	$	&	$	5.55	\pm	1.18	$	&	&	-204.1		&		&	broad	\\	
		&	01-02-25	&	$	20.22	\pm	0.44	$	&	$	6.45	\pm	0.16	$	&	$	13.65	\pm	0.18	$	&	&	-212.0		&		&	broad	\\	
		&	01-02-25	&	$	20.09	\pm	0.43	$	&	$	6.47	\pm	0.16	$	&	$	13.55	\pm	0.17	$	&	&	-214.1	$^{(*)}$	&	-105.7	&	broad	\\	
		&	11-03-25	&	$	5.03	\pm	0.19	$	&	$	1.56	\pm	0.07	$	&	$	2.84	\pm	0.08	$	&	&	-207.0		&		&	broad	\\	\hline

\end{tabular}
\end{table*}

\begin{table*}
\contcaption{}
\centering
\begin{tabular}{lllllllll}
\hline
    &   &  \multicolumn{3}{c}{Flux Measurements} & &\multicolumn{3}{c}{Velocity Measurements}\\
 Star & Date & $F_{2796}$ & $F_{2804,b}$ & $F_{2804,r}$ & & $V_{term}$ & DACs  & Comments  \\
\cline{3-5} \cline{7-9} \\
 & (yy-mm-dd)  &  \multicolumn{3}{c}{($10^{-12} erg\ s^{-1}\ cm^{-2}$)} & &\multicolumn{2}{c}{(km/s)}& \\
\hline
RY	Tau	&	85-03-12	&	$	101.88	\pm	14.62	$	&	$	42.95	\pm	6.87	$	&	$	51.54	\pm	8.15	$	&	&	-267.7		&		&	broad	\\	
		&	85-10-16	&	$	101.87	\pm	14.89	$	&	$	24.76	\pm	5.58	$	&	$	56.61	\pm	6.95	$	&	&	-232.0		&		&	broad	\\	
		&	86-03-22	&	$	97.61	\pm	21.23	$	&	$	9.12	\pm	5.08	$	&	$	67.91	\pm	12.30	$	&	&	-124.9		&		&	broad	\\	
		&	86-10-11	&	$	92.15	\pm	7.09	$	&	$	30.61	\pm	3.92	$	&	$	37.75	\pm	3.22	$	&	&	-142.0		&		&	broad	\\	
		&	87-03-17	&	$	140.75	\pm	16.01	$	&	$	40.08	\pm	6.77	$	&	$	52.71	\pm	6.49	$	&	&	-124.9		&		&	broad	\\	
		&	93-12-31	&	$	86.78	\pm	3.78	$	&	$	27.95	\pm	1.20	$	&	$	32.82	\pm	1.66	$	&	&			&		&	narrow with tail	\\	
		&	01-02-19	&	$	64.32	\pm	1.23	$	&	$	19.15	\pm	0.44	$	&	$	27.58	\pm	0.47	$	&	&	-169.9	$^{(*)}$	&		&	broad and weak	\\	
		&	01-02-20	&	$	64.75	\pm	1.67	$	&	$	19.28	\pm	0.60	$	&	$	27.92	\pm	0.63	$	&	&	-177.7		&	-96.2	&	broad and weak	\\	
		&	01-02-20	&	$	64.14	\pm	1.57	$	&	$	19.44	\pm	0.64	$	&	$	27.93	\pm	0.65	$	&	&	-178.0		&		&	broad and weak	\\	\hline
S	CrA	&	80-05-22	&	$	1.67	\pm	0.75	$	&	$	0.17	\pm	0.33	$	&	$	0.90	\pm	0.39	$	&	&	-311.9		&		&	noisy	\\	\hline
SU	Aur	&	87-10-21	&	$	4.48	\pm	0.76	$	&	$	0.57	\pm	0.22	$	&	$	1.95	\pm	0.39	$	&	&	-123.5		&		&	broad	\\	
		&	87-10-22	&	$	3.52	\pm	0.76	$	&	$	0.50	\pm	0.34	$	&	$	1.63	\pm	0.38	$	&	&	-209.9		&		&	broad	\\	
		&	87-10-23	&	$	3.23	\pm	0.48	$	&	$	0.65	\pm	0.22	$	&	$	2.11	\pm	0.22	$	&	&			&		&		\\	
		&	01-02-24	&	$	2.89	\pm	0.12	$	&	$	0.41	\pm	0.06	$	&	$	1.63	\pm	0.05	$	&	&	-199.9		&	-132.33;-60.7	&	triple structure	\\	
		&	01-02-24	&	$	2.74	\pm	0.11	$	&	$	0.35	\pm	0.06	$	&	$	1.55	\pm	0.04	$	&	&	-195.6		&		&	broad	\\	
		&	11-03-25	&	$	2.76	\pm	0.16	$	&	$	0.45	\pm	0.06	$	&	$	1.49	\pm	0.09	$	&	&	-319.0		&		&	broad	\\	\hline
SZ	102	&	11-05-29	&	$	0.22	\pm	0.02	$	&	$	0.03	\pm	0.01	$	&	$	0.14	\pm	0.02	$	&	&	-200.0		&	-55.6	&	broad	\\	\hline
T	Tau	&	80-11-02	&	$	59.08	\pm	3.90	$	&	$	1.28	\pm	1.19	$	&	$	41.68	\pm	2.15	$	&	&	-295.5		&		&	sharp edge	\\	
		&	80-11-13	&	$	28.36	\pm	4.94	$	&	$	0.23	\pm	0.47	$	&	$	24.39	\pm	2.97	$	&	&	-224.0		&		&	sharp edge	\\	
		&	80-11-14	&	$	29.34	\pm	6.51	$	&	$	2.20	\pm	1.94	$	&	$	36.87	\pm	9.91	$	&	&	-167.0		&		&	sharp edge	\\	
		&	82-03-06	&	$	50.70	\pm	7.31	$	&	$	2.91	\pm	3.32	$	&	$	42.42	\pm	4.66	$	&	&	-158.5		&		&	sharp edge	\\	
		&	95-09-11	&	$	70.01	\pm	1.29	$	&	$	3.63	\pm	0.37	$	&	$	45.03	\pm	0.43	$	&	&	-199.2		&		&	sharp edge	\\	
		&	01-02-21	&	$	40.16	\pm	0.46	$	&	$	1.69	\pm	0.15	$	&	$	25.47	\pm	0.18	$	&	&	-188.5		&		&	sharp edge	\\	
		&	01-02-21	&	$	40.52	\pm	0.43	$	&	$	1.61	\pm	0.13	$	&	$	25.66	\pm	0.16	$	&	&	-188.5	$^{(*)}$	&		&	sharp edge	\\	
		&	01-02-22	&	$	41.07	\pm	0.48	$	&	$	1.75	\pm	0.16	$	&	$	26.16	\pm	0.19	$	&	&	-186.3		&		&	sharp edge	\\	\hline
TW	Hya	&	84-07-16	&	$	2.79	\pm	0.11	$	&	$	0.54	\pm	0.03	$	&	$	1.27	\pm	0.06	$	&	&	-191.3		&		&	narrow with tail	\\	
		&	84-07-16	&	$	1.70	\pm	0.56	$	&	$	0.67	\pm	0.23	$	&	$	0.42	\pm	0.20	$	&	&			&		&	narrow with tail	\\	
		&	00-05-07	&	$	2.33	\pm	0.04	$	&	$	0.52	\pm	0.02	$	&	$	0.99	\pm	0.02	$	&	&	-193.4		&		&	narrow with tail	\\	\hline
TWA	7	&	11-05-05	&	$	0.19	\pm	0.01	$	&	$	0.03	\pm	0.00	$	&	$	0.12	\pm	0.00	$	&	&	-70.0		&		&	narrow with tail	\\	\hline
TWA	3A	&	11-03-26	&	$	0.40	\pm	0.02	$	&	$	0.08	\pm	0.00	$	&	$	0.16	\pm	0.01	$	&	&	-150.0		&		&	narrow with tail	\\	\hline
TWA	13A	&	11-04-02	&	$	0.11	\pm	0.01	$	&	$	0.01	\pm	0.00	$	&	$	0.07	\pm	0.00	$	&	&	-70.0		&		&	narrow with tail	\\	\hline
UX	Tau	&	11-11-10	&	$	0.06	\pm	0.02	$	&	$	0.02	\pm	0.01	$	&	$	0.02	\pm	0.01	$	&	&	-23.0		&		&	narrow absorption	\\	\hline
V819	Tau	&	00-08-31	&	$	0.31	\pm	0.09	$	&	$	0.13	\pm	0.05	$	&	$	0.12	\pm	0.04	$	&	&	-20.0		&		&	narrow absorption	\\	\hline
V836	Tau	&	11-02-05	&	$	0.53	\pm	0.17	$	&	$	0.12	\pm	0.07	$	&	$	0.18	\pm	0.09	$	&	&	-27.0		&		&	narrow absorption	\\	\hline

\end{tabular}
\begin{flushleft}
(*) terminal velocities ($V_{term}$) can be measured accurately.\\
\end{flushleft}

\end{table*}
%%%%%%%%%%%%%%%%%
%%%%%%%%%%%%
With this precaution in mind, we have compared the Mg~II flux with other relevant tracers of the TTSs environment.
We have selected for this purpose:

\begin{itemize}

\item The nearby UV continuum dominated by the Balmer continuum radiation produced in the accretion flow \citep[see e.g.][]{ingleby2013}
measured in the window [2730-2780]~\AA.
\item The C~IV emission produced by hot plasmas in the accretion flow \citep{ardila2013}.
\item The He~II emission from the transition region and/or the accretion shock \citep{aig2013b}.
\item The (P(2)~0-4) 1338.63~\AA\ emission from H$_2$ produced by the molecular gas in the
accretion disc around the TTSs \citep{france2012}. This radiation is, in turn,
pumped by photons at 1217.205~\AA\, in the red wing of the Ly-$\alpha$ line \citep[see e.g.][]{herczeg2002}.

\end{itemize}

\begin{figure*}
\centering
\includegraphics[width=12cm]{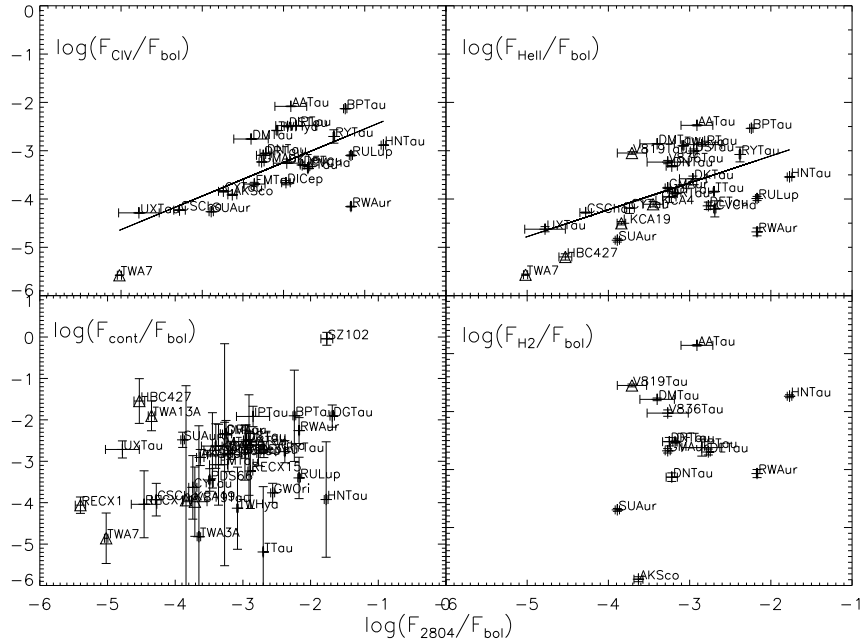}
\caption{Mg~II flux normalized to the bolometric flux of the sources against other relevant UV tracers (see text).
Triangles represent WTTSs. \label{otrosflujos}}
\end{figure*}

As shown in Fig.~\ref{otrosflujos} and in Table~\ref{tablaotrosflujos}, there is a correlation  between the 
Mg~II  and the C~IV flux, as well as with the He~II flux; there is, however, no correlation
with the UV continuum, nor with the H$_2$.

\begin{table}
\caption{Details about the flux-flux correlation between the Mg~II and other spectral tracers.\label{tablaotrosflujos}}
\begin{tabular}{cccc}
\hline
	& No. of & Pearson Cor. & $p{\rm -value}=p^{(*)}$ \\
	& obs. & Coeff.         & 		  \\
\hline

$\frac{F(Mg II)}{F_{bol}}$ vs $\frac{F(UV Cont.)}{F_{bol}}$ & 41 & 0.34 & 0.03 \\
$\frac{F(Mg II)}{F_{bol}}$ vs $\frac{F(C IV)}{F_{bol}}$ & 25 & 0.68 & 0.0002 \\
$\frac{F(Mg II)}{F_{bol}}$ vs $\frac{F(He II)}{F_{bol}}$ & 28 & 0.575& 0.002 \\
$\frac{F(Mg II)}{F_{bol}}$ vs $\frac{F(H_2)}{F_{bol}}$ & 14  & 0.29 & 0.313 \\
\hline
\end{tabular}
\begin{tabular}{l}
$(*)$ $p{\rm -value}=p$ means that, for a random population there is \\
$100 \cdot p$ \% probability that the cross-correlation coefficient \\
will be $r$ or better. We are assuming that the correlation coefficient \\
is statistically significant if the $p{\rm -value}$ is lower than 5\%.
\end{tabular}
\end{table}
%%%%%%%%%%%%%%%%%%
%%%%%%%%%%%%

\subsection{Characterization of the Mg~II profiles: asymmetry, dispersion and broadening}
\begin{figure}
\centering
\includegraphics[width=9cm]{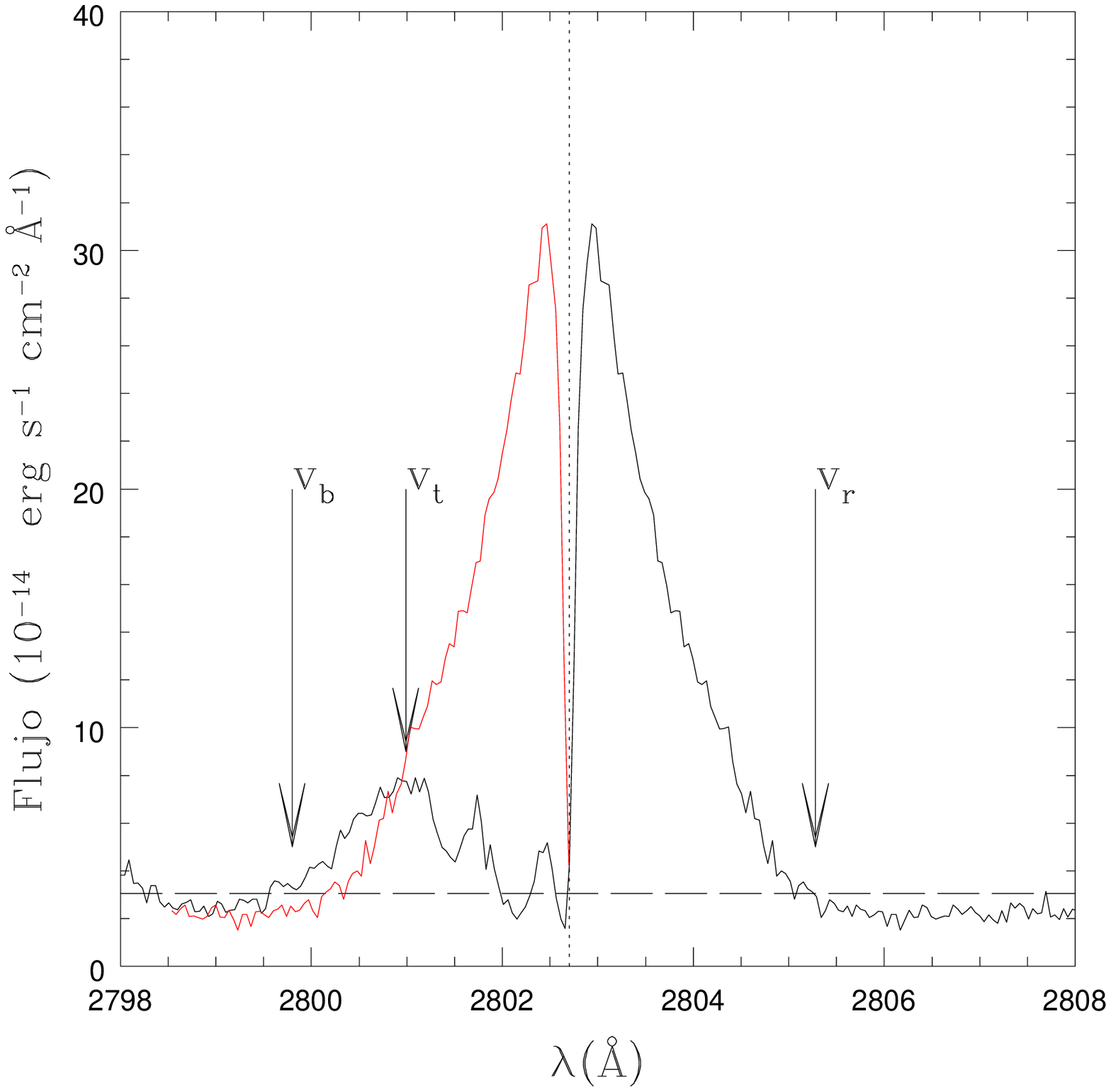} 
\caption{Example profile showing how we measured the terminal velocity of the wind ($V_{term}$) and the blue and red edge velocity of the profile ($V_b$ and $V_r$, respectively). \label{examplevterm}}
\end{figure}

\begin{figure}
\centering
\includegraphics[width=9cm]{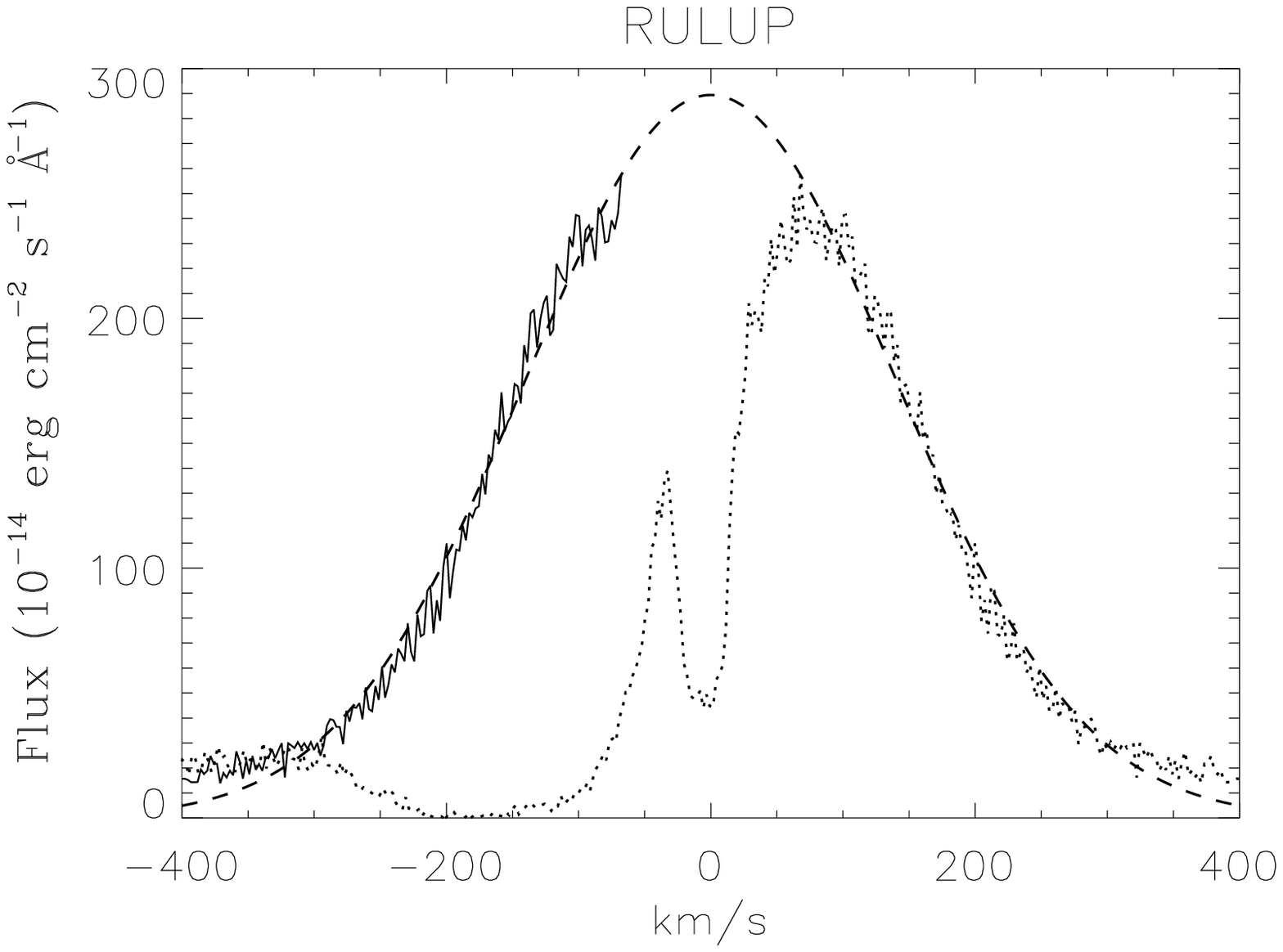}
\caption{The figure illustrates the artificial profile (solid line) used to measure the statistical
moment of the profile plot over the observed line profile (dotted line) and the gaussian fit to the artificial profile (dashed line) for RU Lup.\label{artificalprofile} }
\end{figure}

For this purpose we worked only with the highest S/N profiles for any given source. All measurements
were carried out on the 2804~\AA\ line. We measured:
\begin{itemize}
\item The asymmetry of the profile defined as $A = (F_r)/(F_r+F_b)$, i.e. $A=0.5$ for a symmetric line, 
$A=1$ for a fully absorbed blue wing and $A> 1$ if the absorption goes below the continuum level.  
\item The second statistical moment: dispersion ($\sigma$).
\item The width of the profile at the base of the line $\Delta V$. The red and blue edge velocities of the profile, $V_r$ and $V_b$, are defined as the point where 
the profile meets the continuum plus 1$\sigma$ at the red and blue edges of the profile respectively (see Fig.~\ref{examplevterm}). 
\end{itemize}
%%%%%%%%%%%%%%%%%%%
%%%%%%%%%%%%%%%%%%%%%
Red and blue fluxes ($F_r$ and $F_b$) were measured, as is explained in Section~\ref{fluxes}, from the center of the line to $V_r$ and from $V_b$ to line center, respectively.
The dispersion was calculated from the gaussian fitting to the red wing symmetric profile at 2804~\AA, i.e. 
it is the second statistical moment of the gaussian fit of an artificial and symmetric profile as shown in Fig.~\ref{artificalprofile}. Note that the moment
is calculated avoiding the core of the line and the blue wing, where the circumstellar and the wind 
absorption components reside.
For the stars AK~Sco, CY~Tau, UX~Tau and HBC~388 the dispersion was calculated from the gaussian fit to the
blue wing because the red wing is more absorbed than the blue wing.
%%%%%%%%%%%%%
%%%%%%%%%%%%
\subsubsection{Flux-dispersion relation}
\begin{figure}
\centering
\includegraphics[width=9cm]{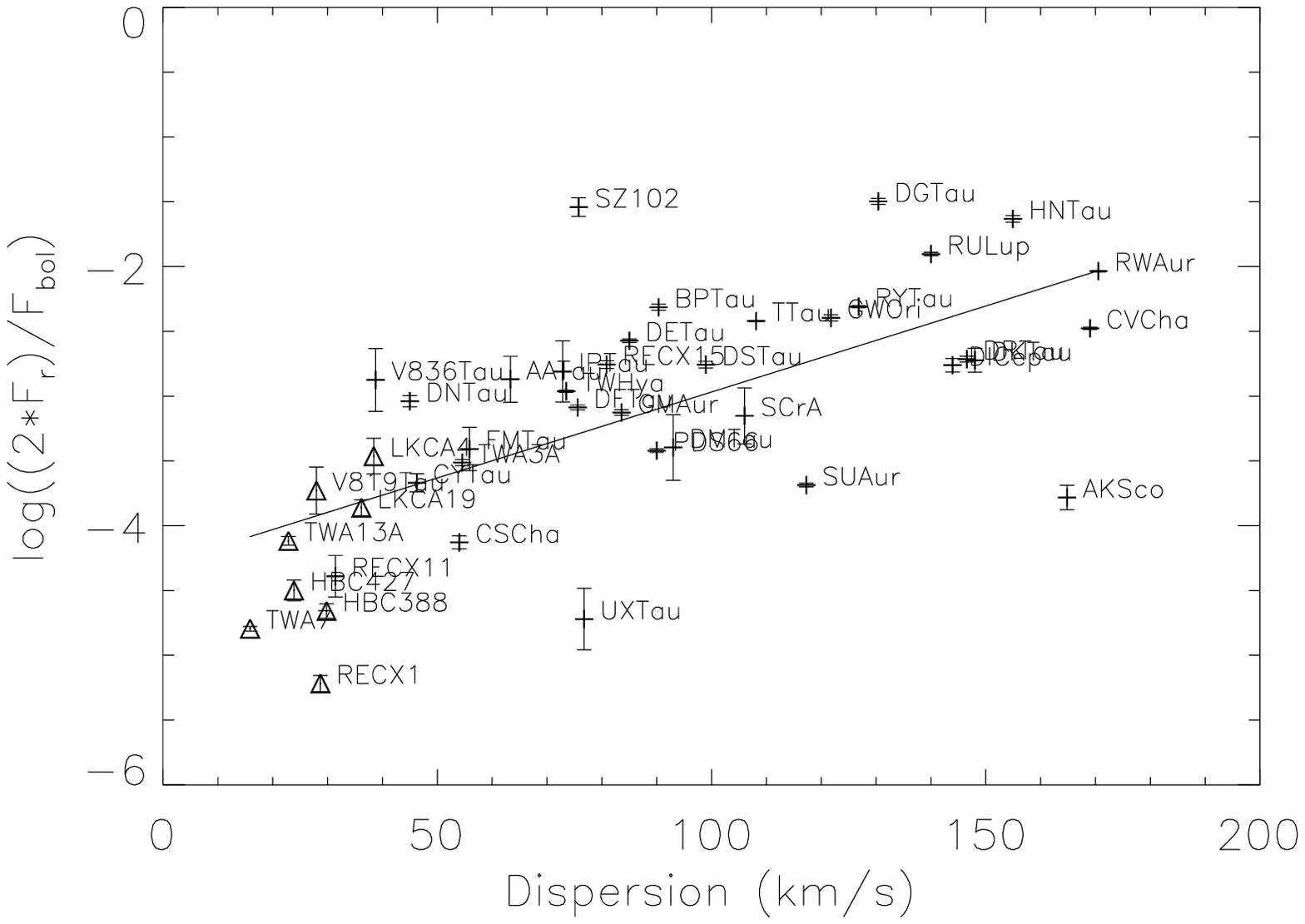}
\includegraphics[width=9cm]{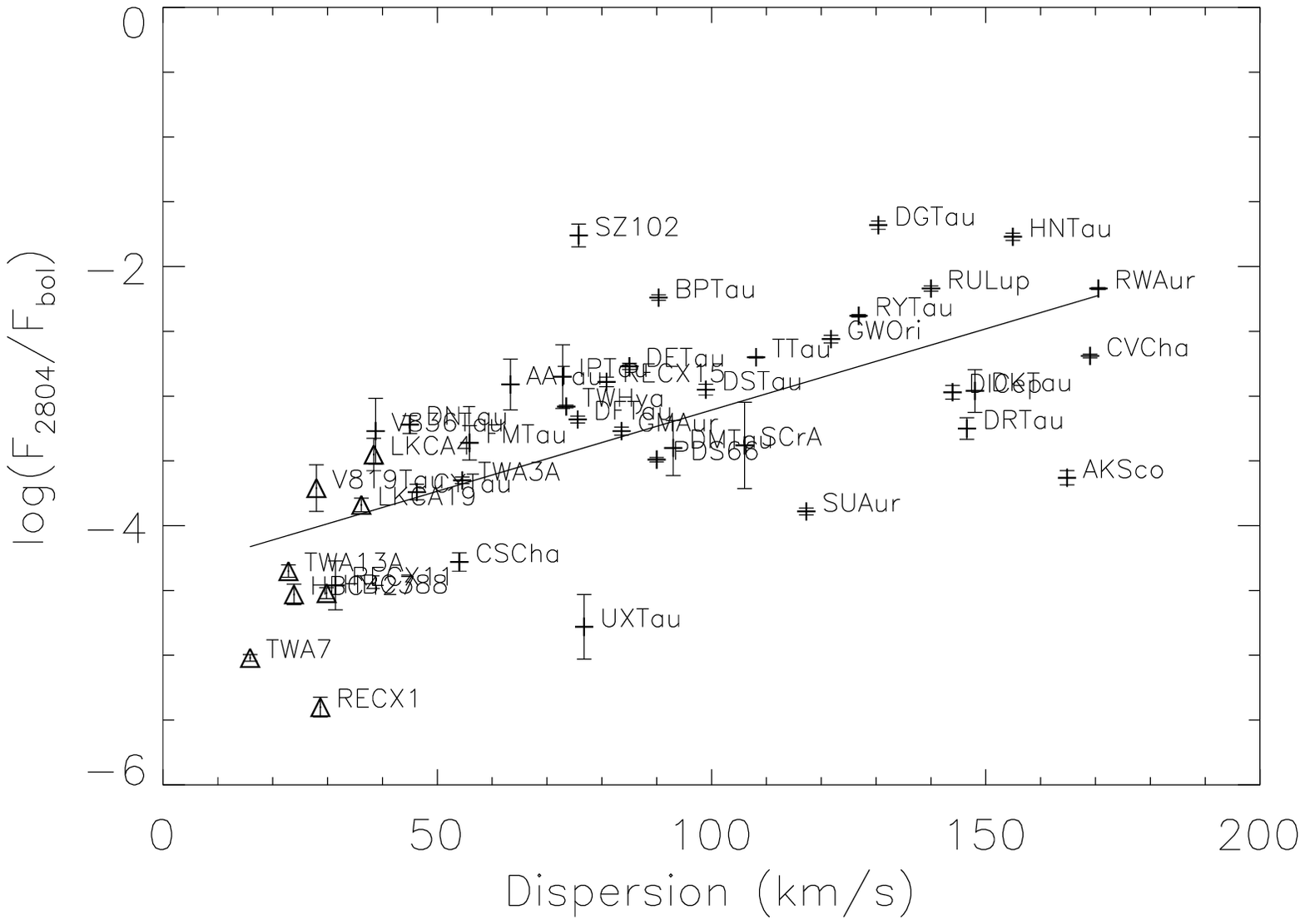}
\caption{Flux-dispersion relation. 
Triangles represent WTTSs. Top: Mg~II flux corrected by effect of the line asymmetry. Bottom: Mg~II flux corrected by effect of luminosity. \label{sigmaflux}}
\end{figure}

WTTSs have small dispersions and CTTSs have larger dispersions.  
The smallest dispersion is $\sigma \simeq 16$~km~s$^{-1}$ for TWA~7.
The highest dispersions are observed for RW~Aur, CV~Cha and AK~Sco with $\sigma \simeq 171$, $169$ and $165$~km~s$^{-1}$, respectively.
For most of WTTSs,
the line broadening is likely due to the plasma thermal velocity ($v_{th} \sim 12$~km~s$^{-1}$
for a $T_e \simeq 10^4$~K)
and the stellar rotation ($v {\rm sin} (i) \simeq 20$~km~s$^{-1}$).
However, line broadening for CTTSs is $50 \la \sigma \la 170$~km~$s^{-1}$, i.e.
larger than the corresponding thermal and rotational velocities.
There is a trend between the Mg~II flux and the line broadening.
Fig.~\ref{sigmaflux} (bottom panel) shows the relation between the Mg~II surface flux  and 
the dispersion with $r=0.66$ and a $p{\rm -value}=1.55 \times 10^{-6}$. 
The correlation does not improve when the profiles are corrected by the wind absorption taking into account the line asymmetry
($r=0.65$ and a $p{\rm -value}=2.86 \times 10^{-6}$),
as shown in the top panel of Fig.~\ref{sigmaflux}.
%%%%%%%%%%%%%%%
%%%%%%%%%%%%%%%%%%
\subsubsection{Dispersion-rotation relation}
\begin{figure}
\centering
\includegraphics[width=9cm]{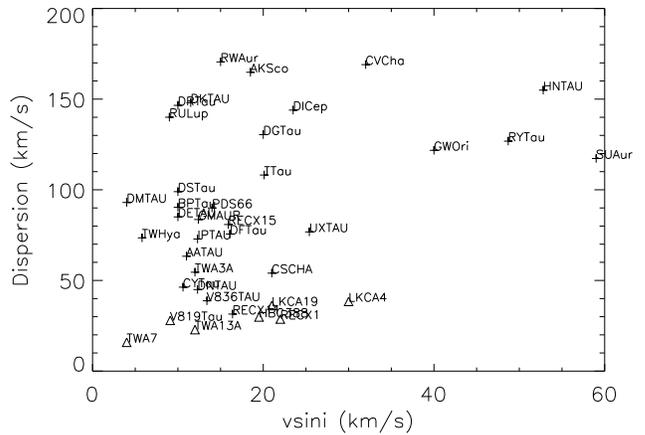}
\caption{Dispersion  as a function of $v{\rm sin}(i)$. 
Triangles represent WTTSs. \label{sigmavrot}}
\end{figure}

To further explore the connection between stellar
rotation $v{\rm sin}(i)$ and line broadening, we plotted them in 
Fig.~\ref{sigmavrot}, 
confirming that the line broadening is not associated with stellar rotation.
%%%%%%%%%%%%%%%%%%
%%%%%%%%%%%
\subsubsection{Flux-asymmetry relation}
\begin{figure}
\centering
\includegraphics[width=9cm]{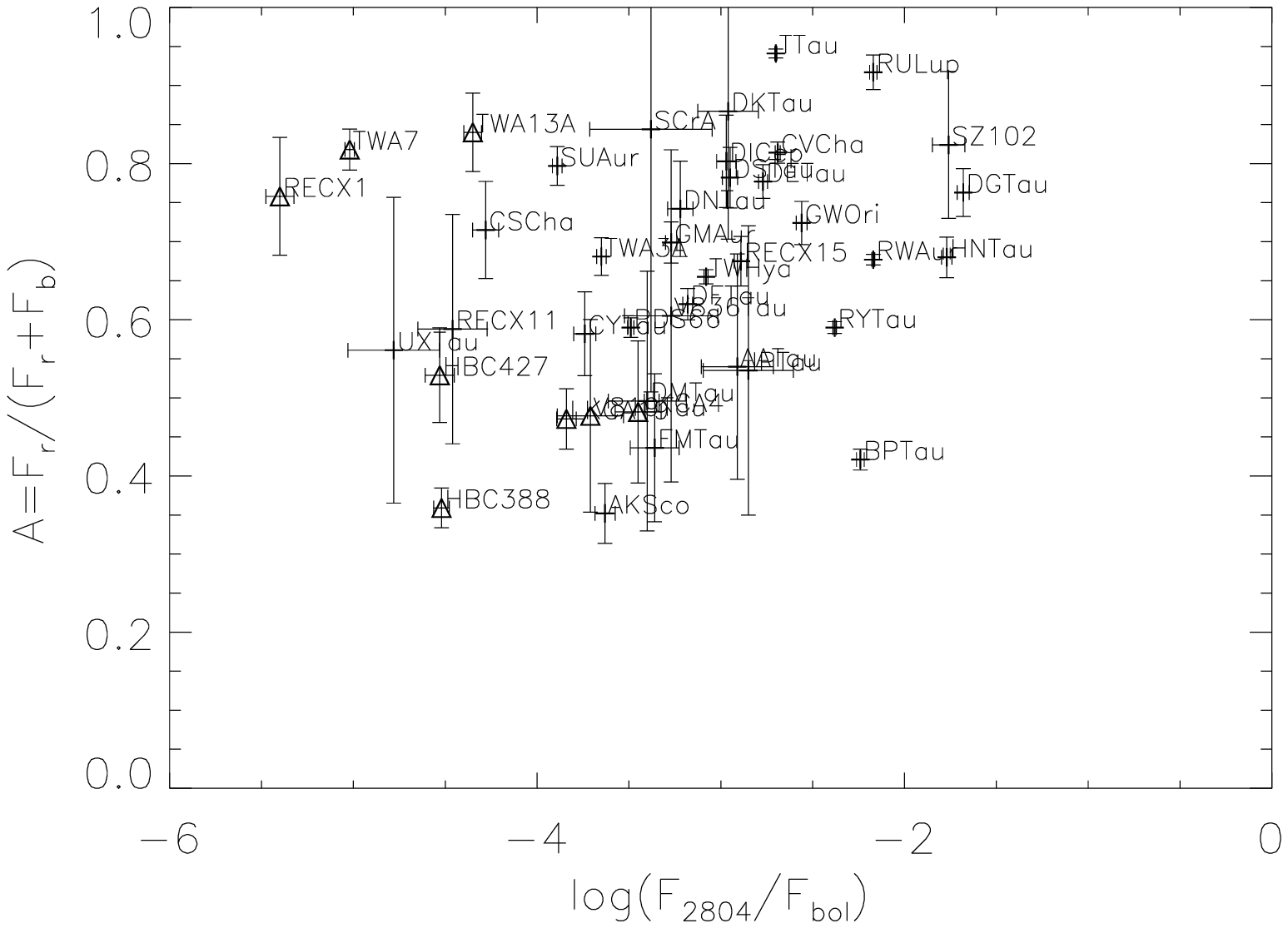}
\caption{Relation between the normalized 2804~\AA\ line asymmetry and 2804~\AA\ line flux. 
Triangles represent WTTSs.\label{asimetriaflux}}
\end{figure}

There is not a correlation between Mg~II flux and profile asymmetry.
This lack of trend is expected if the asymmetry is associated with the orientation
of the outflow with respect to the line of sight (see Fig.~\ref{asimetriaflux}).
WTTSs and CTTSs have similar asymmetry values distributed in a broad range,
from $\sim 0.4$ (for AK~Sco and HBC~388, stars with nearly symmetric profiles)
to $\sim 1$ (for T~Tau, with absorbed blue wing). 
%%%%%%%%%%%%%%%
%%%%%%%%%%%%%%%%%%
\subsubsection{Asymmetry-dispersion-inclination relation}
\begin{figure}
\centering
\includegraphics[width=9cm]{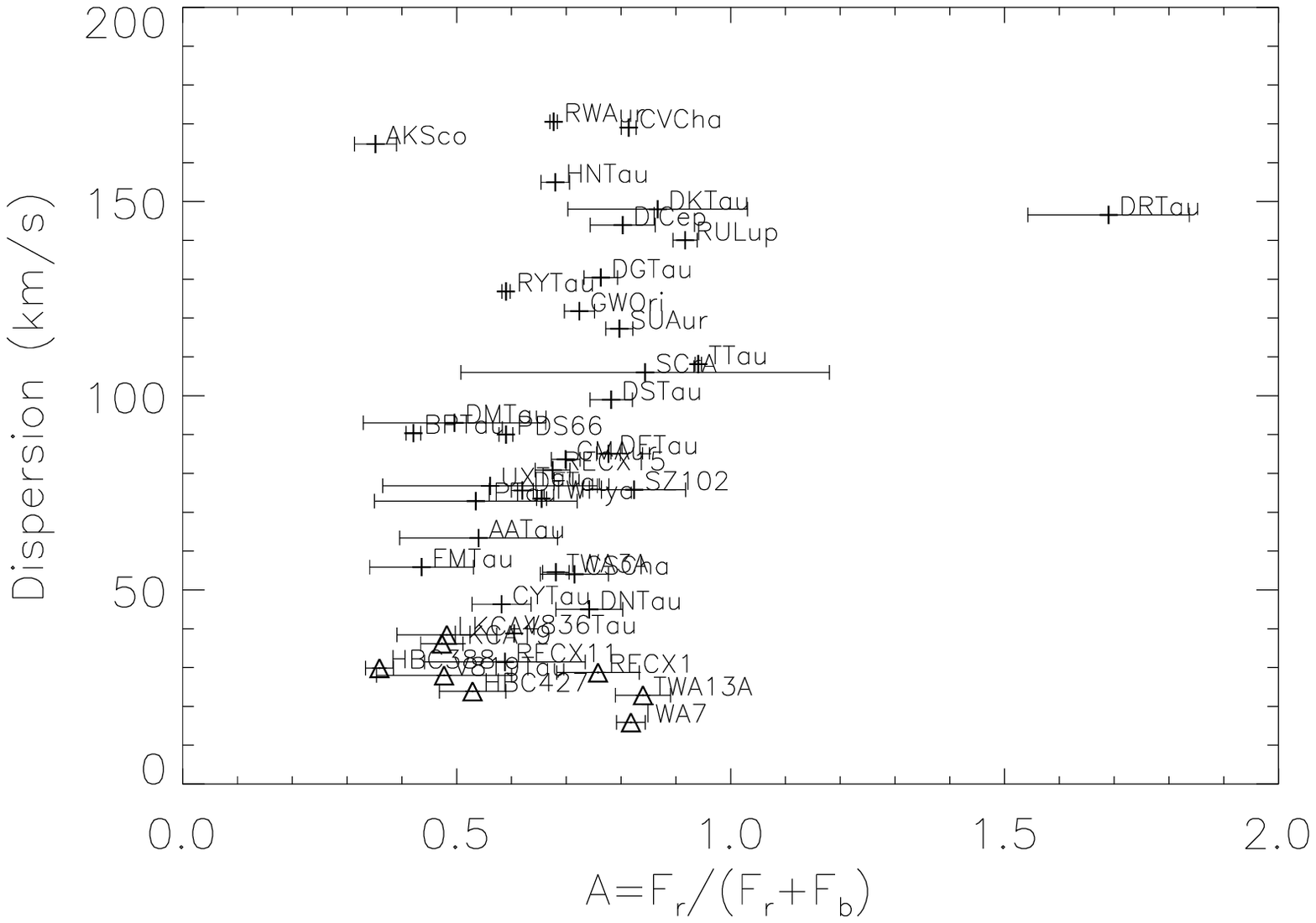} \\
\caption{Relation between dispersion and asymmetry of TTSs in the sample. 
Triangles represent WTTSs. \label{sigmaasimetria}}
\end{figure}

\begin{figure}
\centering
\includegraphics[width=9cm]{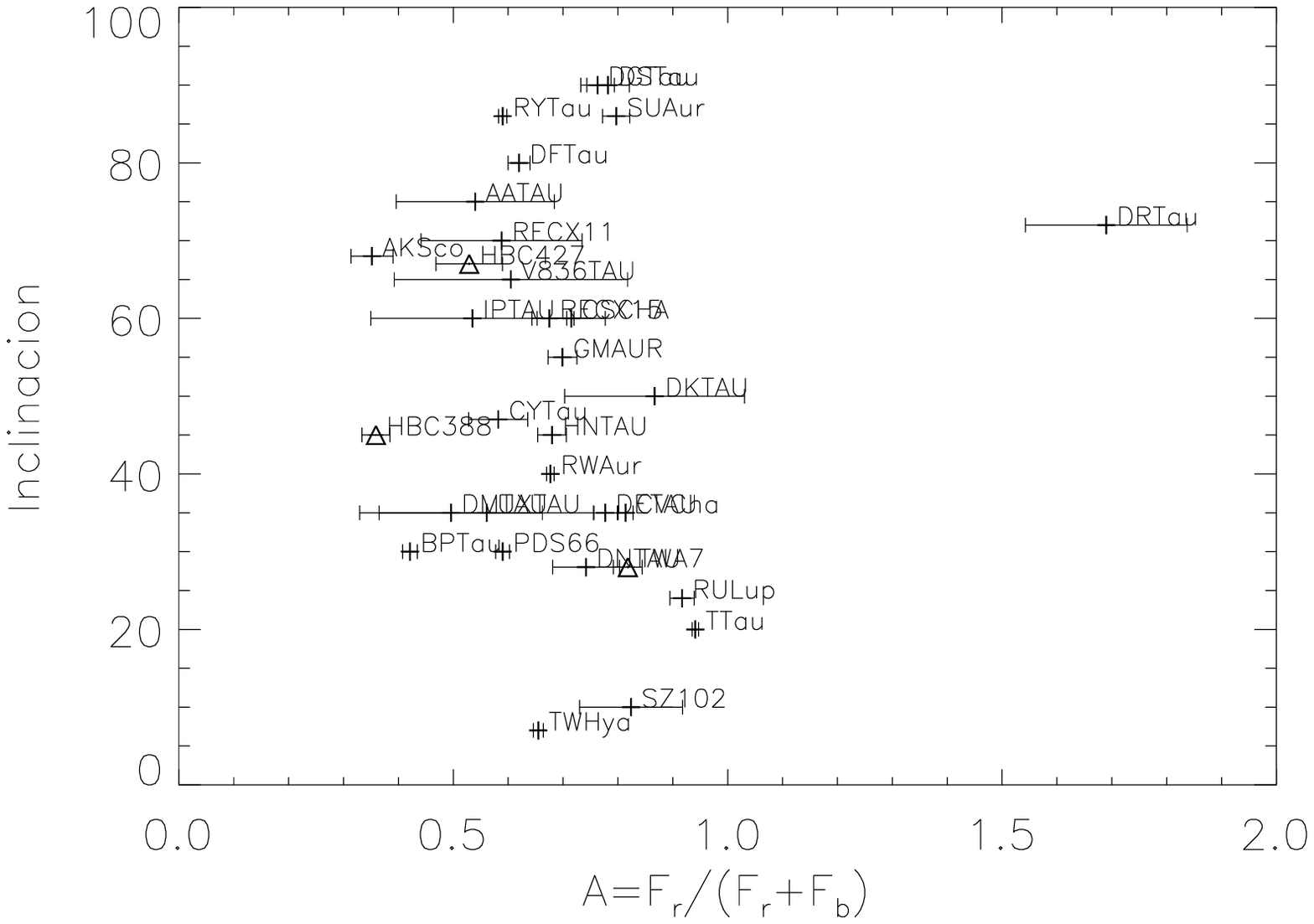}
\caption{Inclination as a function of asymmetry. 
Triangles represent WTTSs. \label{incliasimetria}.}
\end{figure}

Fig.~\ref{sigmaasimetria} shows the dispersion ($\sigma$) as a function of the line asymmetry ($A$), 
where no significant correlation was found.
%%%%%%%%%%%%%%%
%%%%%%%%%%%%
A relation between asymmetry and inclination (see Table~\ref{tabbiblio})
is expected if the asymmetry is associated with the orientation
of the outflow with respect to the line of sight.
However, Fig.~\ref{incliasimetria} does not show this connection.
The lack of correlation could be due to the uncertainties
in inclination measurements. Also, if the outflow is perpendicular to the disc, inclination values derived from stellar $v \sin i$
measurements are not reliable.
%%%%%%%%%%%%
%%%%%%%%%%%%%%
Note that if Mg~II radiation is dominated by the outflow, 
there should be a correlation between inclination and dispersion or
flux that it is not observed.
\subsubsection{Relation between $\Delta V$, $V_r$, $V_b$ and dispersion}
\begin{figure}
\centering
\includegraphics[width=8cm]{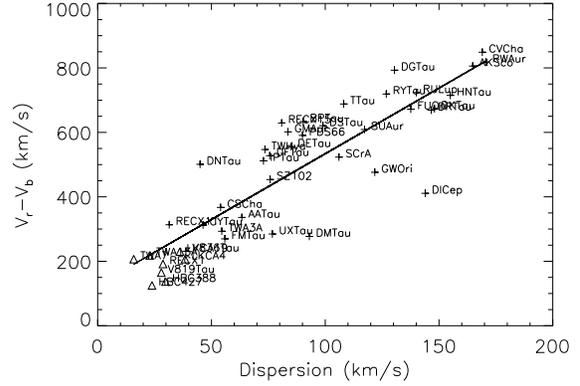}
\includegraphics[width=8cm]{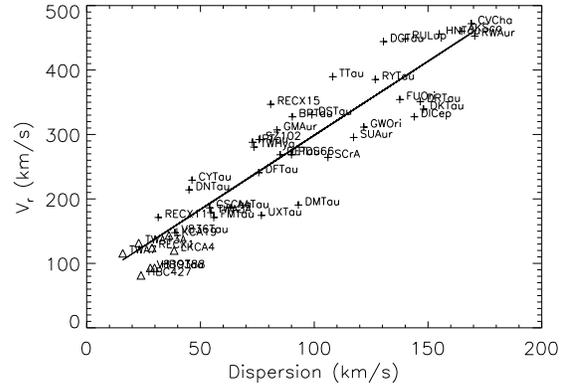}
\includegraphics[width=8cm]{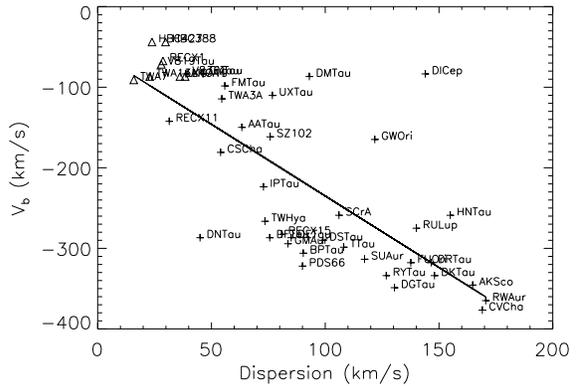}
\caption{Velocities at the profile edge compared with the dispersion. 
Triangles represent WTTSs. \label{sigmavel}}
\end{figure}

\begin{figure}
\centering
\includegraphics[width=9cm]{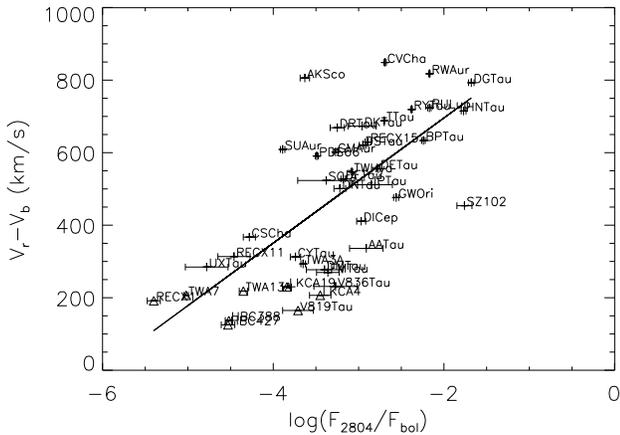}
\caption{Relation between Mg II flux and the Mg II lines width. 
Triangles represent WTTSs. \label{velflux}}
\end{figure}

As shown in Fig.~\ref{sigmavel} (top panel), the dispersion and the width measured at 
the base of the line are correlated ($r=0.9$ and a $p{\rm -value}=0.00$).
In addition the velocities in the blue and red edges of the profile ($V_b$ and $V_r$, respectively) correlate
with the dispersion. This indicates that both line core and the wings radiation are generated by the same phenomenon,
i.e. in the majority of sources the flux in the wings is not dominated by extended features such as unresolved jets.
Otherwise we would observe more extended wings and we would not see this correlation.
%%%%%%%%%%%%%%%
%%%%%%%%%%%%
The correlation in middle panel of Fig.~\ref{sigmavel} between dispersion and $V_r$ is better ($r=0.92$ and a $p{\rm -value}=0.00$)
than with $V_b$ ($r=-0.75$ and a $p{\rm -value}=0.00$) in bottom panel, because of the larger uncertainties
in $V_b$ due to the absorption by the wind.
Moreover, $V_r - V_b$ correlates with the flux,
as we can see in Fig.~\ref{velflux} ($r=0.7$ and
a $p{\rm -value}=1.19 \times 10^{-7}$).
%%%%%%%%%%%%%%%%%%%%%
%%%%%%%%%%%%%%%%%%%%

\subsection{Terminal velocity wind}
Measuring the terminal velocity of the outflow ($V_{term}$) from the Mg~II profiles is challenging in TTSs. There are some sources like DE~Tau or T~Tau that display sharp blue edges and terminal velocities can be measured accurately. These stars are marked in Table~\ref{tabfluxes} with an asterisk. However, for most of the sources, the blue edge is not sharp (see e.g., CV~Cha, FU~Ori or DR~Tau). Moreover, the wind absorption is not observed against the continuum, as in the standard P-Cygni profiles produced by the outflows from massive stars \citep[see for instance][]{talavera1987}. The wind absorption is observed against the broad Mg~II emission from the accretion engine. We have hypothesized that the red wing truly represents the underlying symmetric profile and measured the terminal velocity as the point where the absorption meets the continuum plus $\sigma$ (the 
standard deviation of the continuum), as shown in Fig.~\ref{examplevterm}. The measurements were done independently by the two co-authors and then compared, finding an agreement better than 10~km~s$^{-1}$ between both sets of measurements (see Table~\ref{tabfluxes}, for the results).
For figures, $V_{term}$ corresponding to the best observation (best S/N) is considered. 

Note that the wind structure is very different for the various sources of the sample. In some cases, it just absorbs the blue wing of a 
rather narrow profile (this is typically observed in WTTSs). In other sources, there are double absorption bumps (see e.g. DN~Tau 
or GM~Aur profiles). Unfortunately, there are not time series following the evolution of these components. As shown in the Appendix~\ref{appendix},
the wind absorption (and the profile) do change in the stars for which several observations are available. 

\begin{figure*}
\centering
%\begin{tabular}{ccc}
\includegraphics[width=8cm]{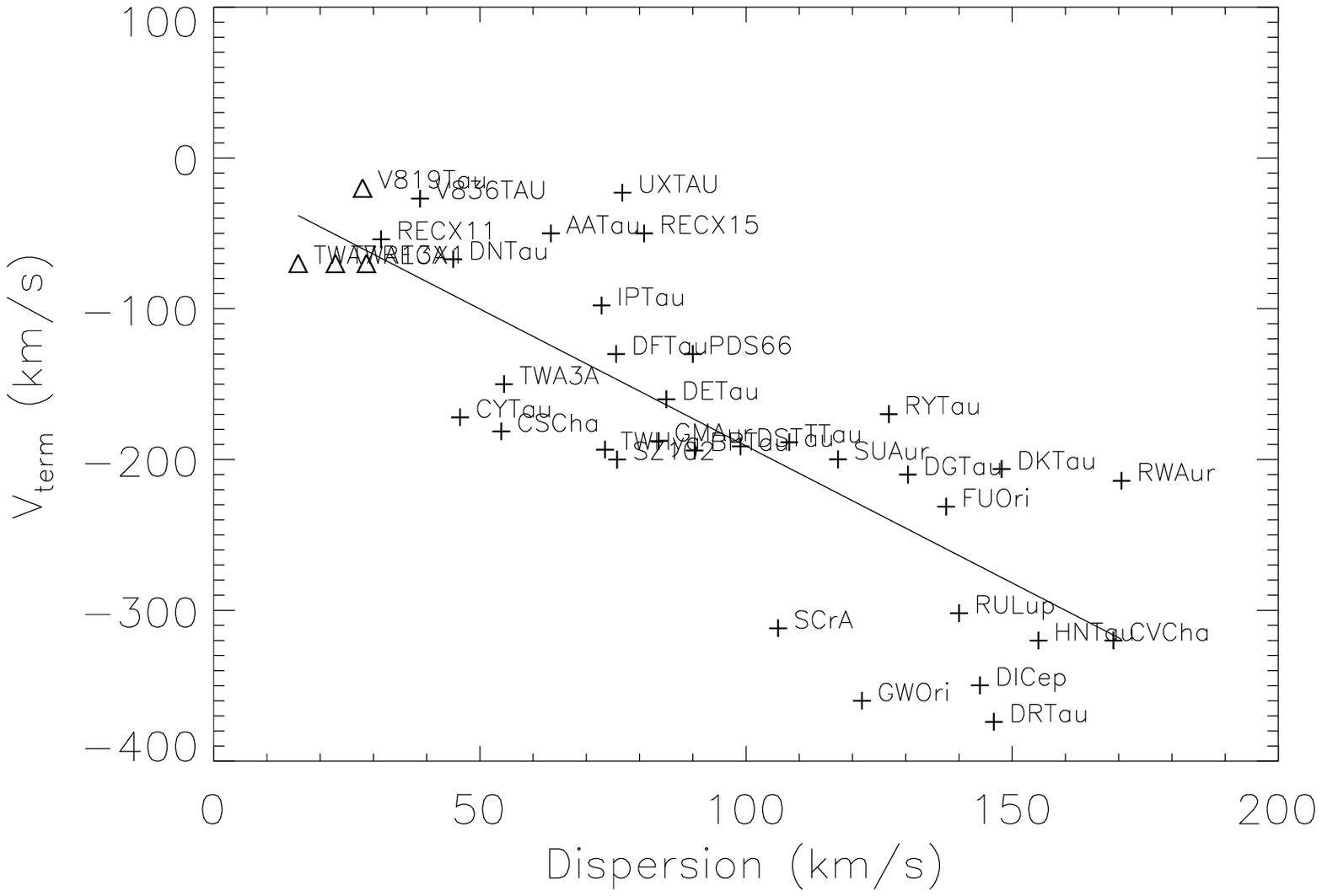}
\includegraphics[width=8cm]{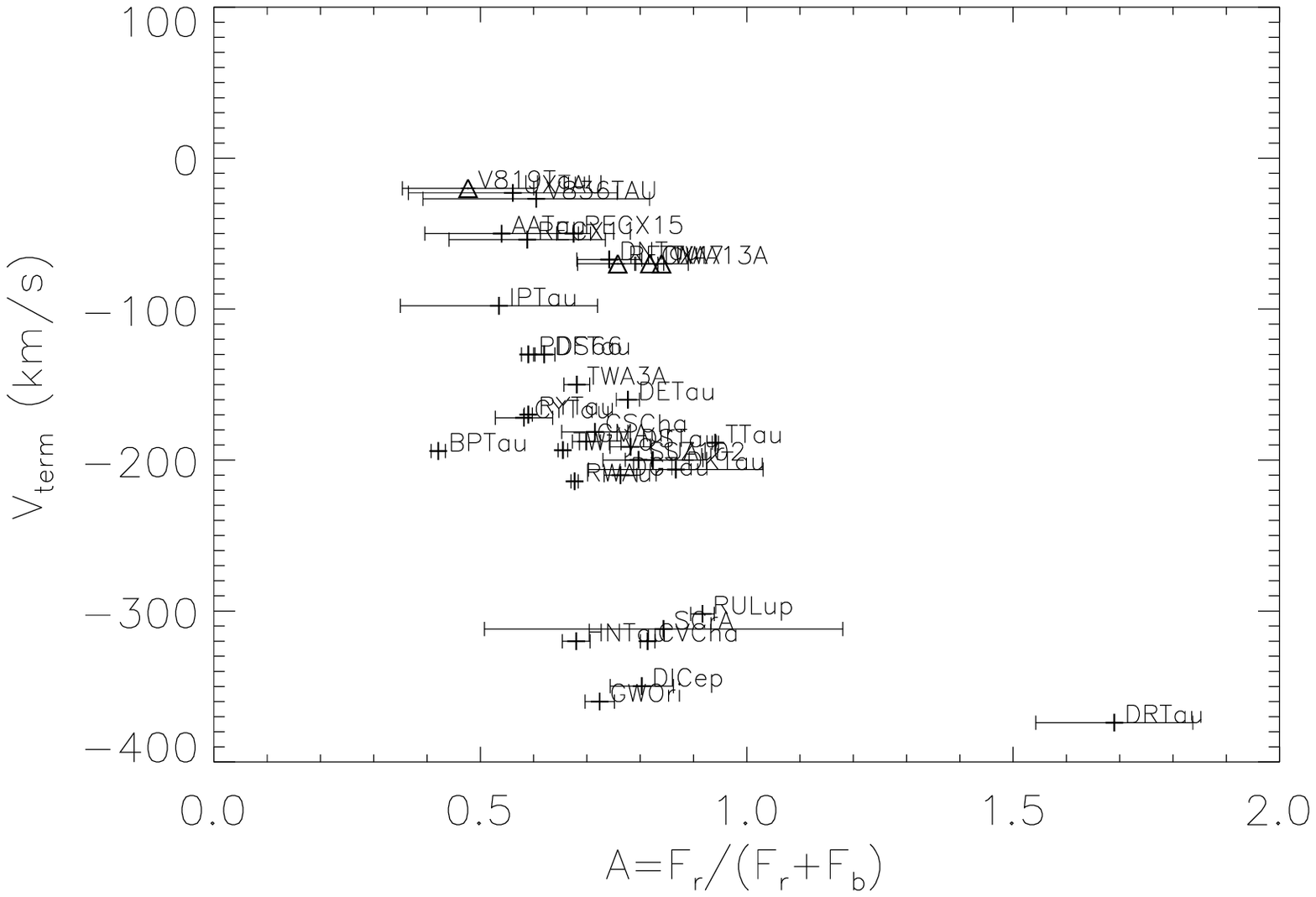}
\includegraphics[width=8cm]{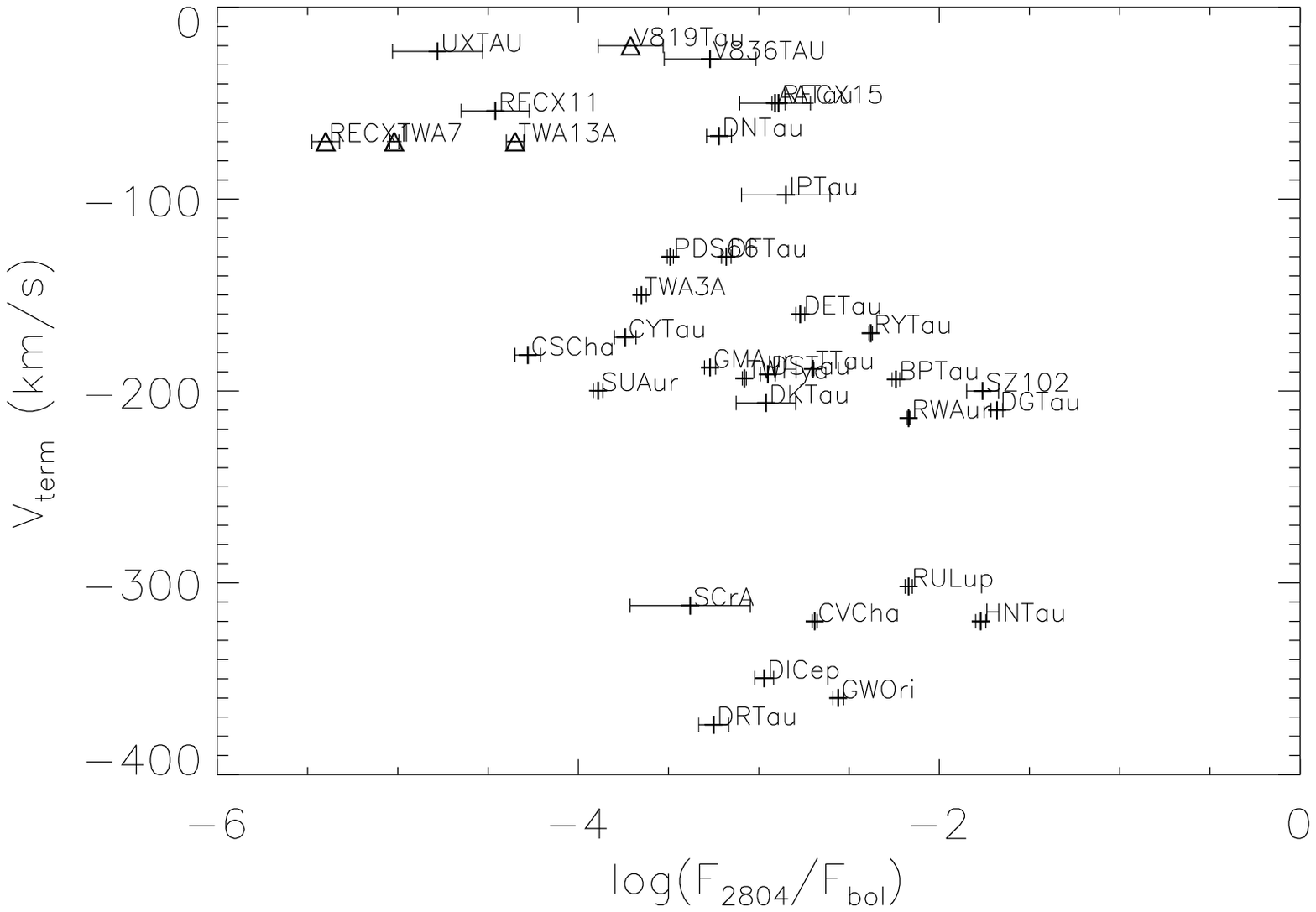} 
%\end{tabular}
\caption{Terminal velocity as a function of dispersion, asymmetry and flux. 
Triangles represent WTTSs. \label{vterm}}
\end{figure*}

In the top left panel of Fig.~\ref{vterm}, the terminal velocity $V_{term}$ it is shown to depend on the profile dispersion
with $r=-0.8$ and a $p{\rm -value}=0.0000$ (for stars with $V_{term}>0$~km~s$^{-1}$).
One may naively think that this correlation is produced by the 
measurement procedure since the absorption is measured against the broad blue-wards shifted emission. However, a careful inspection 
of Fig.~\ref{profiles} shows that $V_{term}$ is controlled by the sharp blue-edge of the wind absorption. 
We did not find a significant correlation between the terminal velocity neither with profile asymmetry nor with flux (see top right and bottom panel of Fig.~\ref{vterm}).
%%%%%%%%%%%%%%
%%%%%%%%%%%%%%%%
\subsection{Profile variability}
Repeated observations are available for 17 stars but significant profile variations are observed only in some of them, namely, BP~Tau, RY~Tau, T~Tau, DF~Tau, DG~Tau, DR~Tau, RW~Aur, TW~Hya and RU~Lup (see Table~\ref{tabfluxes} and Appendix~\ref{appendix}). In most of them, the variations are associated with the absorption components in the blue wing of the profile and they are more noticeable in profiles displaying several absorption components than in those displaying broad, saturated absorption components; blue wing absorptions seem to be associated to variable or episodic  ejection. The terminal velocity of the flow varies only in few sources: DR~Tau, DS~Tau, FU~Or, RU~Lup, RY~Tau, SU~Aur and T~Tau.

\section{Constraints to the physics of TTS outflows}
\label{constraints}
This work provides some important constraints to the physics of the TTSs outflows.
The first constraint derives from the comparison between the Ly-$\alpha$ and Mg~II profiles. 
As shown in Fig.~\ref{profiles}, the blueshifted absorption produced by the wind,
the sharp blue-edge indicating that the terminal velocity is reached,
the variable discrete absorption components observed in some sources (e.g., SU~Aur
and BP~Tau) are observed in the Mg~II lines and remain undetectable 
in the Ly-$\alpha$ profile, even in the unabsorbed wings.
This indicates that the {\it wind is  warm} ($\log {\rm T}_{\rm e} \simeq 4-4.3$) 
{\it and keeps a rather constant temperature in the acceleration region}.
Note that the absorption ranges from small velocities to typical protostellar jets speeds.
This temperature regime is cooler than the detected in the semiforbidden Si~III], C~III] transitions \citep{aig2001,aig2007}. 

The Mg~II profiles show that the wind covers a broad range of projected velocities along the line of sight.
This, in turn, indicates that either the wind is kept isothermal while expanding radially or the outflow geometry 
is not radial, even at the base of the wind. Since large scale outflows from TTSs are collimated, this observation sets-up scales of several stellar radii for wind collimation.

The detection of DACs in some sources suggests that mass ejection 
is episodic even at small scales. This is consistent
with the observations of knots in optical jets. 
One may wonder whether the broad absorptions are caused by the blending of many DACs, at least in some sources.
Note that the current framework for modelling of mass ejection in TTSs includes episodic phenomenon produced by reconnection events in the magnetospheric star-disc boundary layer, see e.g. \citet*{rekowski2004,rekowski2006}, earlier works by \citet{goodson1997,goodson1999,goodson21999} or later works by \citet{romanova2012}. 
However, Mg~II observations show smooth absorption profiles. Henceforth, either the environmental conditions 
in early phases are such that the density in the current layer makes it prone to more frequent reconnection 
(for instance the higher density) producing a blending of broad DACs showing as a smooth absorption profile or Mg~II 
is tracing another wind component, more likely the disc wind. 
Some constraints to the physics of TTS outflows are:

{\it Evidence of latitude-dependent outflow on stellar scales from the profile asymmetry}. 
A larger asymmetry is expected for pole-on systems, since the wind produces a larger blue wing
absorption in these systems.
However, we did not find a connection between inclination and profile asymmetries.
This lack of correlation could be due to the uncertainties
of the inclinations.
In addition, we studied the possible relations among
asymmetry and other magnitudes, such as flux, dispersion and terminal velocity of the wind,
but we did not find any significant correlation.

{\it Relation between Mg~II emission and wind/outflow.}
The processes responsible for line broadening are related with Mg~II flux emission (see Fig.\ref{sigmaflux}).
Fig.~\ref{vterm} (top left panel) shows a connection between
dispersion and terminal velocity of the wind.
However, it is unclear from the data whether this connection is direct,
i.e., the outflow contributes significantly to the Mg~II line emission,
or indirect through the well reported connection between accretion rate and wind
signatures \citep[see, for instance,][]{cabrit1990,aig1993}.
In a recent work \citep{fatima2014}, we have shown that the radiation in
single ionized forbidden lines such as C~II], Fe~II] and Si~II]
is dominated by the extended stellar magnetosphere and the accretion flow.
These lines trace a thermal regime similar to that traced by the
Mg~II lines, thus though we might expect a contribution from
the wind to the Mg~II flux, it seems accreting plasma dominates the line radiation.
Therefore, the correlation we report between the dispersion and the 
$v_{term}$ is most likely indirect, both observables depend on the accretion rate.

\begin{figure}
\centering
\includegraphics[width=9cm]{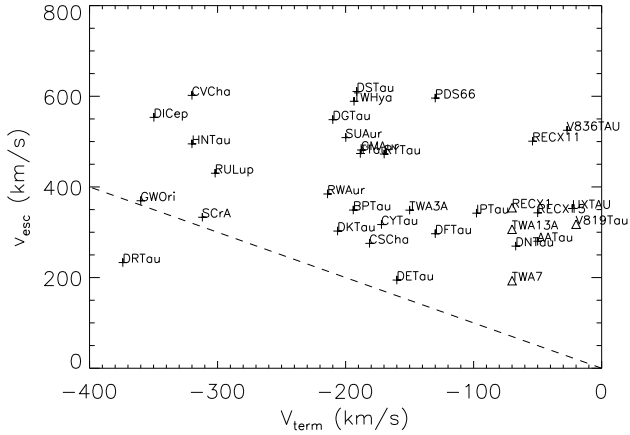}
\caption{Escape velocity ($v_{esc}$) from stellar surface as a function of the Terminal velocity ($v_{term}$). Dashed line indicates where $v_{term}=v_{esc}$.
Triangles represent WTTSs. \label{f18}}
\end{figure}

{\it The role of the gravitational field in mass ejection}. 
To study where wind launching occurs, we have examined the relation between terminal velocity of the flow and escape velocity ($V_{esc}$) from the stellar surface (see Fig~\ref{f18}). Escape velocity was computed as $v_{esc} = \sqrt{2GM_*/R_*} $ where $R_*=\sqrt{L_*/ (4 \pi \sigma T_{eff}^4)} $ using $L_*$, M$_*$ and T$_{eff}$ from Table~\ref{tabbiblio}.
The escape velocities from stellar surface are larger than
the terminal velocities of wind measured in the profiles.
All measurements satisfy $V_{esc} < V_{term}$ going from the
small $V_{esc}= -150$~km/s to the high $V_{esc} = -350$~km/s 
values. This suggests there is some scaling law between
escape and projected terminal velocity.
Note that our $V_{term}$ values
correspond to the projection of the terminal velocity in the line
of sight and thus, it is affected by inclination effects.   

%%%%%%%%%%%%%%%%%%%%
%%%%%%%%%%%%%%%%%%%
\section{Relation between the Mg II radiation and magnetospheric emission}
\label{discussion}
%flujo y mdot
The temperature of the TTSs magnetospheres is expected to be rather cool, about some few
thousands Kelvin \citep{romanova2012,kulkarni2013}. Hence, we might expect that an 
uncertain fraction of the line flux is produced in the magnetosphere and that magnetospheric rotation
and turbulence produce the line broadening.
 
\begin{figure}
\centering
\includegraphics[width=9cm]{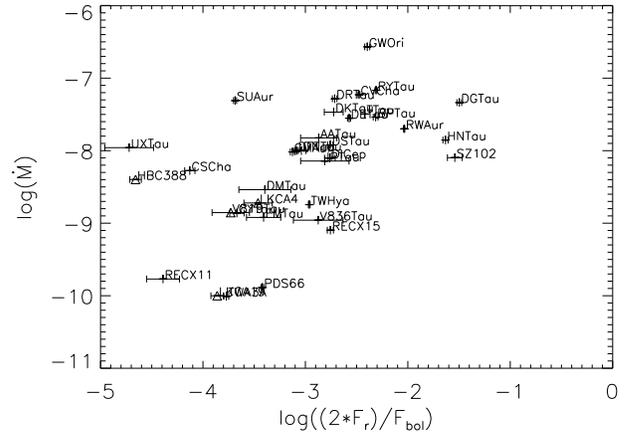}
\includegraphics[width=9cm]{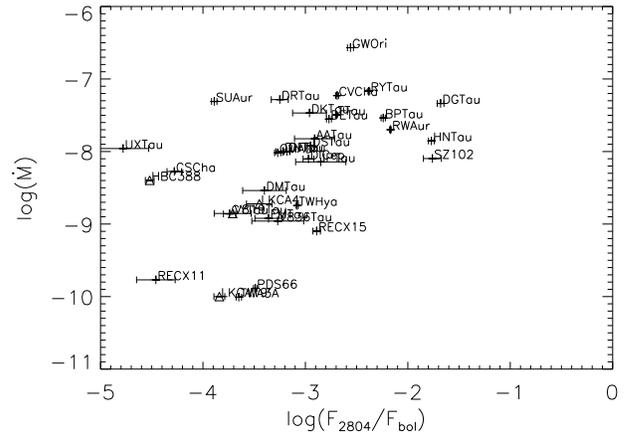}
\caption{Comparison between accretion rate ($M_{\sun} yr^{-1}$) and Mg II flux (2804~\AA). 
Triangles represent WTTSs. Top: Mg~II flux corrected by
asymmetry and luminosity effects. Bottom: Mg~II corrected by luminosity effect. \label{mdotflux}}
\end{figure}

\begin{figure}
\centering
\includegraphics[width=9cm]{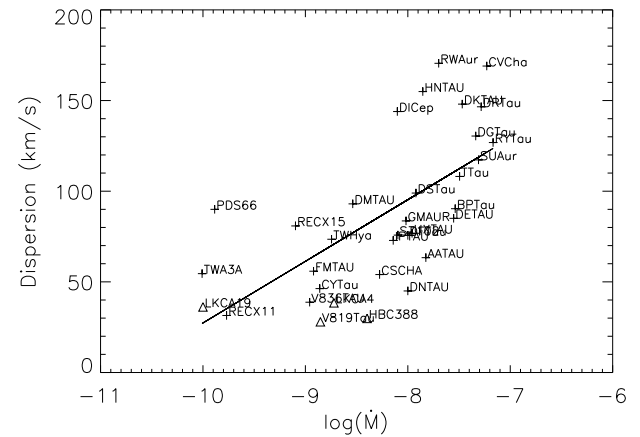}
\caption{Relation between the dispersion and the accretion rate ($M_{\sun} yr^{-1}$). 
Triangles represent WTTSs.\label{sigmamdot}}
\end{figure}

\begin{figure}
\centering
\includegraphics[width=9cm]{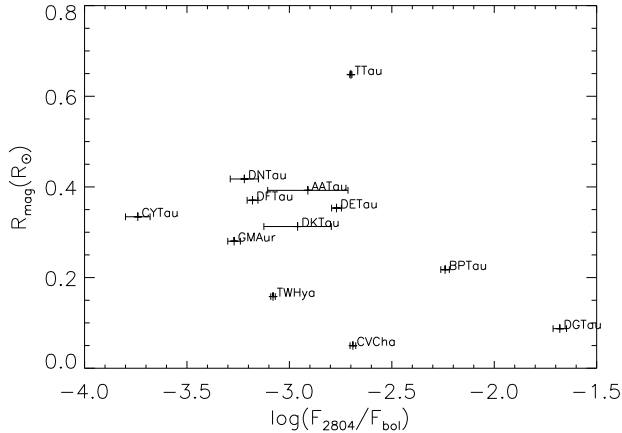}
\caption{Relation between normalized 2804~\AA\ line flux and magnetospheric radius ($R_{mag}$). \label{fluxrmag}}
\end{figure}

In this context, it is confusing the lack of a strong correlation between accretion rate and Mg~II flux,
see the bottom panel of Fig.~\ref{mdotflux} ($r=0.5$ and $p{\rm -value}=0.002$) and also Fig.~\ref{otrosflujos} for the
correlation with the UV continuum radiation.
Correcting the Mg~II from the wind absorption does not improve significantly the trend; $r=0.54$
and a $p{\rm -value}=0.0007$ (see top panel of Fig.~\ref{mdotflux}). 
Hence, radiation from the accretion flow does not seem to dominate the bulk of the Mg~II radiation.
We note that accretion rates used in this work come
form other authors measurements \citep[mainly from][]{ingleby2013}
and thus Mg~II and accretion rate measurements are not simultaneous.
As shown in Fig.~\ref{observations}, fluxes can vary by a factor of 2 in
accreting sources and this variability may affect to the reported lack
of correlation.
%%%%%%%%%%%%%%%%%%%%%%%%
%%%%%%%%%%%%%%%%%%%%%%
However, there seems to be a correlation between the line broadening and the accretion rate, as shown in
Fig.~\ref{sigmamdot} ($r=0.66$ and $p{\rm -value}=2.14 \times 10^{-5}$).
%%%%%%%%%%%%
%%%%%%%%%%%%
%flujo y radio magnetico
We searched a connection between the strength of the Mg~II emission and the magnetospheric radius, that could provide some hints on the role of magnetospheric radiation on the dissipation of the angular momentum excess \citep[see][for a recent study]{aig2012}. \\
The size of the magnetosphere is set by the balance between the toroidal component of the stellar magnetic flux and the angular momentum of the infalling matter \citep{ghosh1979}: 
$$
R_{\rm mag} = \left(\frac {\gamma ^2 \mu ^4}{G M_* M_{acc}^2}\right)^{1/7} 
$$
with $\mu = B_* R_*^3$, the equatorial magnetic moment of the star, $\gamma = (B_t/B_p)(\Delta r /r) \simeq 0.5-0.8$ \citep[see][]{lamb1989}, $M_{acc}$ is the accretion rate and $B_*$ is the surface magnetic field.
The magnetospheric radius can be calculated for a small subset of the stars. 
Surface magnetic fields were measured for AA~Tau, DE~Tau, DK~Tau, DN~Tau, GM~Aur, T~Tau, CY~Tau, BP~Tau, DF~Tau, DG~Tau, TW~Hya \citep[see][]{johnskrull2007} and CV~Cha \citep{gregory2012}.
However, no significant relation was found between the radius in this way determined and the Mg~II normalized flux (see Fig.~\ref{fluxrmag}). 
Thus, not conclusive results can be inferred from the plot.\\
%%%%%%%%%%%%
%%%%%%%%%%%%
\section{Conclusions}
\label{conclusions}
The analysis of the TTS Mg~II profiles has provided new insights on the behaviour of the wind engine, including the magnetosphere, the accretion flow and the outflow.
The main conclusions that have been drawn from this work are:

\begin{enumerate}
\item There is a warm wind that at sub-AU scales absorbs the blue wing of the Mg~II profile. Thus,
these lines are an ideal tracer of the wind acceleration region. 

\item We find a relation between the line broadening both with
the terminal velocity and with the accretion rate.
This result could be an evidence that the accretion drives the winds/outflows processes.

\item The profile broadening, as measured from the dispersion, correlates with the velocity at the edge of the wings ($V_r$, $V_b$)
\item A mild correlation is found between the Mg~II flux and the accretion rate. 

\item We find a connection between line broadening and Mg~II flux.
Both outflow and magnetospheric plasma contribute to the Mg~II flux; however,
separating both contributions is very complex and model-dependent, monitoring programs are
needed for this type of work.
\end{enumerate}
 
We would like to emphasize the current uncertainties in age and mass for PMS stars.
This work shows the potentials of high resolution UV spectroscopy to study the wind engine in PMS stars. 
Dedicated monitoring programs would be fundamental to study the wind acceleration region,
especially in sources such as RY~Tau where variable discrete components have been detected.
 
\section*{acknowledgements}

The authors acknowledge support from the Spanish Ministry of Economy and Competitiveness through grant BES-2009-014629 associated to investigation project World Space Observatory-Ultraviolet (WSO-UV): AYA2008-06423-C03-01 and AYA2011-29754-C03-01. 
Fatima L\'opez-Mart\'inez is grateful to Nestor Sanchez for his useful comments. 
Ana I. G\'omez de Castro thanks Kevin France for his comments on the COS observations and Suzanne Edwards, Greg Herczeg, Sergey Lamzin and Jeff Linsky, for interesting conversations about TTSs physics.
We also wish to thank an anonymous referee for her/his useful comments.

\appendix
\section{Log of the Ly-$\alpha$ observations and variability of the Mg~II profiles}
\label{appendix}
In this section we include the log of the Ly-$\alpha$ observations (see Table~\ref{lymanalpha}).
Also, the figures showing the variability of the Mg~II profiles in the TTSs (see Fig. that is available online) are shown. Only observations with good enough S/N are compared.

%%%%%%%%%%%%%%%%%%%%%%%%%%%%
\begin{table*}
\caption{Telescope/instrument details for the Ly-$\alpha$ observations of the stars in the sample. The full table is available online as Supporting Information.\label{lymanalpha}}
\begin{tabular}{cccccc}
\hline
Star & Instrument & Obs. Date & Data set & Res. &  Exposure \\
     &            & (yy-mm-dd) & Id.     &   power         &  Time (s)    \\ \hline
AA	Tau	&	HST/COS	&	11-01-06		&	LB6B07040	&	19000	&	2844.3	\\	
		&	HST/COS	&	11-01-07		&	LB6B07050	&	18000	&	2844.4	\\	\hline
AK	Sco	&	HST/STIS	&	10-08-21	&	OB6B21040	&	45800	&	2917.2	\\	\hline
BP	Tau	&	HST/COS	&	11-09-09		&	LBGJ01050	&	18000	&	1150.2	\\	
		&	HST/COS	&	11-09-09		&	LBGJ01060	&	18000	&	1190.2	\\	
		&	HST/COS	&	11-09-09		&	LBGJ01070	&	19000	&	1397.2	\\	
		&	HST/COS	&	11-09-09		&	LBGJ01080	&	19000	&	1396.2	\\	\hline
CS	Cha	&	HST/COS	&	11-06-01		&	LB6B16010	&	18000	&	1177.9	\\	
		&	HST/COS	&	11-06-01		&	LB6B16020	&	19000	&	1177.9	\\	\hline
CV	Cha	&	HST/STIS	&	13-11-04		&	OB6B18030	&	45800	&	3265.2	\\	\hline
\end{tabular}
\end{table*}
%%%%%%%%%%%%%%%%%%%%%%%%

\end{document}